\documentstyle[editedvolume,numreferences,epsf]{crckapb}

\newcommand{\be}{\begin{equation}}
\newcommand{\ee}{\end{equation}}
\newcommand{\la}{\langle}
\newcommand{\ra}{\rangle}

\begin{opening}
\title{INFLATIONARY COSMOLOGY: PROGRESS AND PROBLEMS 
}
 
\author{ROBERT H. BRANDENBERGER }
\institute{Physics Department, Brown University\\
           Providence, RI, 02912, USA}
 
\end{opening}

\runningtitle{INFLATIONARY COSMOLOGY}

\begin{document}

\begin{abstract}
These lecture notes {\footnote{Brown preprint BROWN-HET-1196, invited lectures at the International School on Cosmology, Kish Island, Iran, Jan. 22 - Feb. 4 1999, to be publ. in the proceedings (Kluwer, Dordrecht, 2000)}} intend to form a short pedagogical introduction to inflationary cosmology, highlighting selected areas of recent progress 
such as reheating and the theory of cosmological perturbations. Problems of principle for inflationary cosmology are pointed out, and some new attempts at solving them are indicated, including a nonsingular Universe construction by means of higher derivative terms in the gravitational action, and the study of back-reaction of cosmological perturbations.
\end{abstract}

\section{Introduction}

Inflationary cosmology \cite{Guth} has become one of the cornerstones of modern cosmology. Inflation was the first theory which made predictions about the structure of the Universe on large scales based on causal physics. The development of the inflationary Universe scenario has opened up a new and extremely promising avenue for connecting fundamental physics with experiment.

These lectures form a short pedagogical introduction to inflation, focusing more on the basic principles than on detailed particle physics model building. Section 2 outlines some of the basic problems of standard cosmology which served as a motivation for the development of inflationary cosmology, especially the apparent impossibility of having a causal theory of structure formation within the context of standard cosmology. 

In Section 3, it is shown how the basic idea of inflation can solve the horizon and flatness problems, and can lead to a causal theory of structure formation. It is shown that when trying to implement the idea of inflation, one is automatically driven to consider the interplay between particle physics / field theory and cosmology. The section ends with a brief survey of some models of inflation. 

Section 4 reviews two areas in which there has been major progress since the early days of inflation. The first topic is reheating. It is shown that parametric resonance effects may play a crucial role in reheating the Universe at the end of inflation. The second topic is the quantum theory of the generation and evolution of cosmological perturbations which has become a cornerstone for precision calculations of observable quantities.

In spite of the remarkable success of the inflationary Universe paradigm, there are several serious problems of principle for current models of inflation, specifically potential-driven models. These problems are discussed in Section 5.

Section 6 is a summary of some new approaches to solving the problems of potential-driven inflation. An attempt to obtain inflation from condensates is discussed, a nonsingular Universe construction making use of higher derivative terms in the gravitational action is explained, and a framework for calculating the back-reaction of cosmological perturbations is summarized.

As indicated before, this review focuses on the principles and problems of inflationary cosmology. Readers interested in comprehensive reviews of inflation are referred to \cite{Linde,GuthBlau,Olive}. A recent review focusing on inflationary model building in the context of supersymmetric models can be found in \cite{LR99}. For a review at a similar level to this one but with a different bias see \cite{Liddle2}.
 
\section{Successes and Problems of Standard Cosmology}

\subsection{Framework of Standard Cosmology}

The standard big bang cosmology rests on three theoretical pillars: the
cosmological principle, Einstein's general theory of relativity and a classical perfect fluid description of matter.

The cosmological principle states that on large distance scales the
Universe is homogeneous. This implies that the metric of space-time can be written in Friedmann-Robertson-Walker (FRW) form:
\be
 ds^2 = dt^2 - a(t)^2 \, \left[ {dr^2\over{1-kr^2}} + r^2 (d \vartheta^2 + \sin^2 \vartheta d\varphi^2) \right] \, , 
\ee
where the constant $k$ determines the topology of the spatial sections. In the following, we shall usually set $k = 0$, i.e. consider a spatially flat Universe. In this case, we can set the scale factor $a(t)$ to be equal to $1$ at the present time $t_0$, i.e. $a(t_0) = 1$, without loss of generality. The coordinates $r, \vartheta$ and $\varphi$ are comoving spherical coordinates. World lines with constant comoving coordinates are geodesics corresponding to particles at rest. If the Universe is expanding, i.e. $a(t)$ is increasing, then the physical distance $\Delta x_p(t)$ between two points at rest with fixed comoving distance $\Delta x_c$ grows:
\be
\Delta x_p = a(t) \Delta x_c \, . 
\ee
 
The dynamics of an expanding Universe is determined by the Einstein equations,
which relate the expansion  rate to the matter content, specifically to the
energy density $\rho$ and pressure $p$.  For a homogeneous and isotropic
Universe and setting the cosmological constant to zero, they reduce to the Friedmann-Robertston-Walker (FRW) equations
\be
\left( {\dot a \over a} \right)^2 - {k\over a^2} = {8 \pi G\over 3 } \rho
\ee
\be
{\ddot a\over a} = - {4 \pi G\over 3} \, (\rho + 3 p) \, .
\ee
These equations can be combined to yield the continuity equation (with Hubble
constant $H = \dot a/a$)
\be \label{cont}
\dot \rho = - 3 H (\rho + p) \, . 
\ee

The third key assumption of standard cosmology is that matter is described by
a classical ideal gas with an equation of state
\be
p = w \rho \, . 
\ee
For cold matter ({\it dust}), pressure is negligible and hence $w = 0$.  From (\ref{cont}) it
follows that
\be
\rho_m (t) \sim a^{-3} (t) \, , 
\ee
where $\rho_m$ is the energy density in cold matter.  For radiation we have $w
= {1/3}$ and hence it follows from (\ref{cont}) that
\be
\rho_r (t) \sim a^{-4} (t) \, , 
\ee
$\rho_r (t)$ being the energy density in radiation.
 
\subsection{Successes of Standard Cosmology}     

The three classic observational pillars of standard cosmology are Hubble's law, the existence and black body nature of the nearly isotropic cosmic microwave background (CMB), and the abundances of light elements (nucleosynthesis). These successes are discussed in detail in many textbooks (see e.g. \cite{Peebles,Padmanabhan,Peacock} for some recent ones) on cosmology, and also in the lectures by Blanchard and Sarkar at this school, and will therefore not be reviewed here.

Let us just recall two important aspects of the thermal history of the early Universe. Since the energy density in radiation redshifts faster than the matter energy density, it follows that although the energy density of the Universe is now mostly in cold matter, it was initially dominated by radiation. The transition occurred at a time denoted by $t_{eq}$, the ``time of equal matter and radiation". As discussed in the lectures by Padmanabhan at this school, $t_{eq}$ is the time when perturbations can start to grow by gravitational clustering. The second important time is $t_{rec}$, the ``time of recombination" when photons fell out of equilibrium (since ions and electrons had by then combined to form electrically neutral atoms). The photons of the CMB have travelled without scattering from $t_{rec}$ to the present. Their spatial distribution is predicted to be a black body since the cosmological redshift preserves the black body nature of the initial spectrum (simply redshifting the temperature) which was in turn determined by thermal equilibrium. CMB anisotropies probe the density fluctuations at $t_{rec}$ (see the lectures by Zadellariaga at this school for a detailed analysis). Note that for the usual values of the cosmological parameters, $t_{eq} < t_{rec}$. 

\subsection{Problems of Standard Cosmology}

Standard Big Bang cosmology is faced with several important problems.  None represents an actual conflict with observations.  The
problems I will focus on here -- the homogeneity, flatness and formation of
structure problems (see e.g. \cite{Guth}) -- are questions which have no answer
within the theory and are therefore the main motivation for inflationary cosmology.

The ``horizon problem" is illustrated in Fig. 1.  As is sketched, the comoving
region $\ell_p (t_{rec})$ over which the CMB is observed to be homogeneous  to
better  than one part in $10^4$ is much larger than the comoving forward light
cone $\ell_f (t_{rec})$ at $t_{rec}$, which is the maximal distance over which
microphysical forces could have caused the homogeneity:
\be
\ell_p (t_{rec}) = \int\limits^{t_0}_{t_{rec}} dt \, a^{-1} (t) \simeq 3 \, t_0
\left(1 - \left({t_{rec}\over t_0} \right)^{1/3} \right) 
\ee
\be
\ell_f (t_{rec}) = \int\limits^{t_{rec}}_0 dt \, a^{-1} (t) \simeq 3 \, t^{2/3}_0
\, t^{1/3}_{rec} \, . 
\ee
{}From the above equations it is obvious that $\ell_p (t_{rec}) \gg \ell_f
(t_{rec})$.  Hence, standard cosmology cannot explain the observed isotropy of
the CMB.

\begin{figure} 
\begin{center}
\leavevmode
\epsfysize=6.5cm \epsfbox{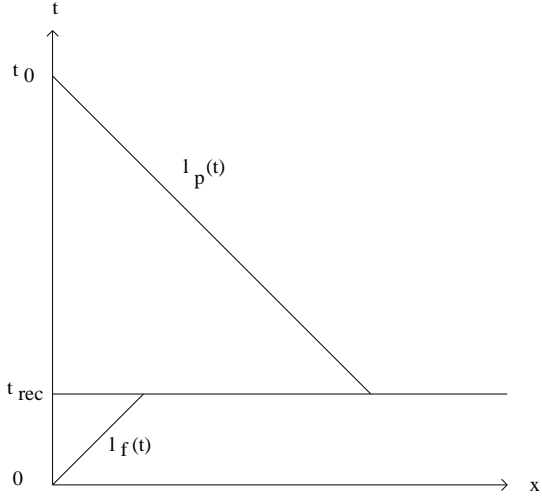}
\caption{
A space-time diagram (physical distance $x_p$ versus time $t$)
illustrating the homogeneity problem: the past light cone $\ell_p (t)$ at the
time $t_{rec}$ of last scattering  is much larger than the forward light cone
$\ell_f (t)$ at $t_{rec}$.}
\end{center}
\end{figure}

In standard cosmology and in an expanding Universe with conserved total entropy, $\Omega = 1$ is an unstable
fixed point.  This can be seen as follows.  For a spatially flat Universe
$(\Omega = 1)$
\be \label{omega1}
H^2 = {8 \pi G\over 3} \, \rho_c \, , 
\ee
whereas for a nonflat Universe
\be \label{omega2}
H^2 + \varepsilon \, T^2 = {8 \pi G\over 3}  \, \rho \, , 
\ee
with
\be
\varepsilon = {k\over{(aT)^2}} \, . 
\ee
The quantity $\varepsilon$ is proportional to $s^{-2/3}$, where $s$ is the
comoving entropy density.  Hence, in standard cosmology, $\varepsilon$ is constant.
Combining (\ref{omega1}) and (\ref{omega2}) gives
\be \label{omega3}
{\rho - \rho_c\over \rho_c} = {3\over{8 \pi G}} \, {\varepsilon T^2\over
\rho_c} \sim T^{-2} \, . 
\ee
Thus, as the temperature decreases, $\vert \Omega - 1 \vert$ increases.  In fact, in order
to explain the present small value of $\Omega  \simeq 1$, the
initial energy density had to be extremely close to critical density.  For
example, at $T = 10^{15}$ GeV, (\ref{omega3}) implies
\be
{\rho - \rho_c\over \rho_c} \sim 10^{-50} \, . 
\ee
What is the origin of these fine tuned initial conditions?  This is the
``flatness problem" of standard cosmology.

The third of the classic problems of standard cosmological model is the
``formation of structure problem."  Observations indicate that galaxies and
even clusters of galaxies have nonrandom correlations on scales larger than 50
Mpc (see e.g. \cite{CFA,LCRS}).  This scale is comparable to the comoving horizon at
$t_{eq}$.  Thus, if the initial density perturbations were produced much before
$t_{eq}$, the correlations cannot be explained by a causal mechanism.  Gravity
alone is, in general, too weak to build up correlations on the scale of
clusters after $t_{eq}$ (see, however, the explosion scenario of \cite{explosion} and topological defect models \cite{VilShell,HK95,RB94} discussed in these proceedings in \cite{MB99}).
Hence, the two questions of what generates the primordial density perturbations
and what causes the observed correlations do not have an answer in the context
of standard cosmology.  This problem is illustrated by Fig. 2.

\begin{figure}
\begin{center}
\leavevmode
\epsfysize=7.5cm \epsfbox{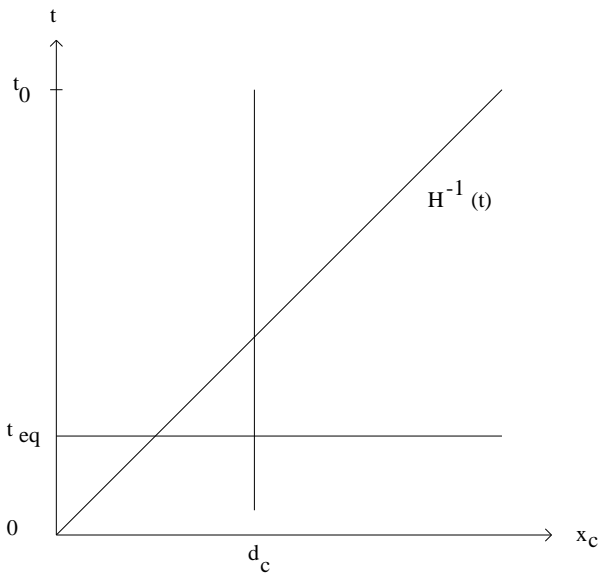}
\caption{A sketch (conformal separation vs. time) of the formation of
structure problem: the comoving separation $d_c$ between two clusters is larger
than the forward light cone at time $t_{eq}$.}
\end{center}
\end{figure}

Standard cosmology extrapolated all the way back to the big bang cannot be taken as a self-consistent theory. The theory predicts that as the big bang is approached the temperature of matter diverges. This implies that the classical ideal gas description of matter which is one of the pillars of the theory breaks down. This comment serves as a guide to which of the key assumptions of standard cosmology will have to be replaced in order to obtain an improved theory: this improved theory will have to be based on the best theory available which describes matter at high temperatures and energies. Currently the best available matter theory is quantum field theory. In the near future, however, quantum field theory may have to be replaced by the theory which extends it to even higher energies, most likely string theory.
 
\section{Overview of Inflationary Cosmology}
 
\subsection{The Inflationary Scenario}

The idea of inflation \cite{Guth} is very simple (for some early reviews of inflation see e.g. \cite{Linde,GuthBlau,Olive,RB85}).  We assume there is a time
interval beginning at $t_i$ and ending at $t_R$ (the ``reheating time") during
which the Universe is exponentially expanding, i.e.,
\be
a (t) \sim e^{Ht}, \>\>\>\>\> t \epsilon \, [ t_i , \, t_R] 
\ee
with constant Hubble expansion parameter $H$.  Such a period is called  ``de
Sitter" or ``inflationary."  The success of Big Bang nucleosynthesis sets an
upper limit to the time $t_R$ of reheating:
\be
t_R \ll t_{NS} \, , 
\ee
$t_{NS}$ being the time of nucleosynthesis.

\begin{figure}
\begin{center}
\leavevmode
\epsfxsize=12cm \epsfbox{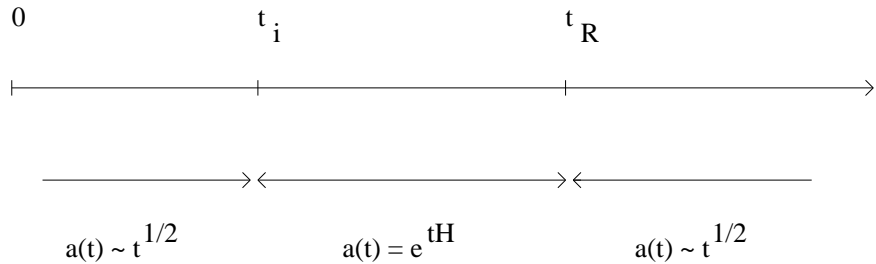}
\caption{The time line of an inflationary Universe. The times
$t_i$ and $t_R$ denote the beginning and end of inflation, respectively.
In some models of inflation, there is no initial radiation dominated FRW
period. Rather, the classical space-time emerges directly in an inflationary
state from some initial quantum gravity state.}
\end{center}
\end{figure}  

The phases of an inflationary Universe are sketched in Fig. 3.  Before the
onset of inflation there are no constraints on the state of the Universe.  In
some models a classical space-time emerges immediately in an inflationary
state, in others there is an initial radiation dominated FRW period.  Our
sketch applies to the second case.  After $t_R$, the Universe is very hot and
dense, and the subsequent evolution is as in standard cosmology.  During the
inflationary phase, the number density of any particles initially in thermal
equilibrium at $t = t_i$ decays exponentially.  Hence, the matter temperature
$T_m (t)$ also decays exponentially.  At $t = t_R$, all of the energy which is
responsible for inflation (see later) is released as thermal energy.  This is a
nonadiabatic process during which the entropy increases by a large factor.   

Fig. 4 is a sketch of how a period of inflation can solve the homogeneity
problem.  $\Delta t = t_R - t_i$  is the period of inflation.  During
inflation, the forward light cone increases exponentially compared to a model
without inflation, whereas the past light cone is not affected for $t \geq
t_R$.  Hence, provided $\Delta t$ is sufficiently large, $\ell_f (t_R)$ will be
greater than $\ell_p (t_R)$. 

\begin{figure}
\begin{center}
\leavevmode
\epsfysize=9cm \epsfbox{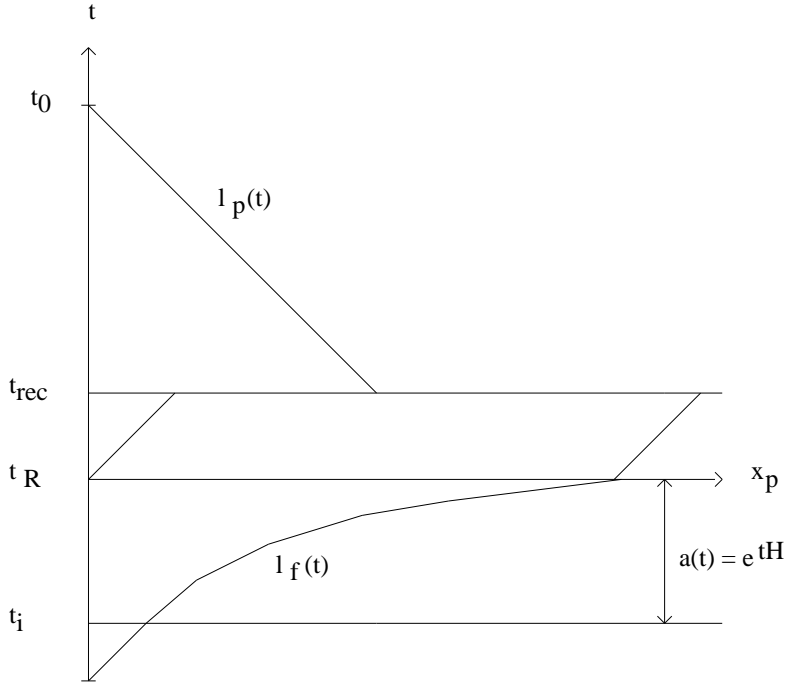}
\caption{
Sketch (physical coordinates vs. time) of the
solution of the homogeneity problem. During inflation, the forward light cone
$l_f(t)$ is expanded exponentially when measured in physical coordinates.
Hence, it does not require many e-foldings of inflation in order that $l_f(t)$
becomes larger than the past light cone at the time of last scattering. The
dashed line is the forward light cone without inflation.}
\end{center}
\end{figure}  

Inflation also can solve the flatness problem \cite{Kazanas,Guth}.  The key point is
that the entropy density $s$ is no longer constant.  As will be explained
later, the temperatures at $t_i$ and $t_R$ are essentially equal.  Hence, the
entropy increases during inflation by a factor $\exp (3 H \Delta t)$.  Thus,
$\epsilon$ decreases by a factor of $\exp (-2 H \Delta t)$.  Hence, $\rho$ and $\rho_c$ can be of comparable magnitude at both $t_i$ and the
present time.  In fact, if inflation occurs at all, then rather generically, the theory predicts
that at the present time $\Omega = 1$ to a high accuracy (now $\Omega < 1$
requires  special initial conditions or rather special models \cite{open}).

Most importantly, inflation provides a causal mechanism for
generating the primordial perturbations required for galaxies, clusters and even
larger objects.  In inflationary Universe models, the Hubble radius
(``apparent" horizon), $3t$, and the (``actual") horizon (the forward light cone)
do not coincide at late times.  Provided that the duration of inflation is sufficiently long, then (as sketched
in Fig. 5) all scales within our present apparent horizon were inside the
horizon since $t_i$.  Thus, in principle it is possible to have a casual
generation mechanism for perturbations \cite{Press,Mukh80,Lukash,Sato}.

The generation of perturbations is supposed to be due to a causal microphysical
process.  Such processes can only act coherently on length scales smaller than
the Hubble radius $\ell_H (t)$, where
\be
\ell_H (t) = H^{-1} (t) \, . 
\ee
A heuristic way to understand $\ell_H (t)$ is to realize that it
is the distance which light (and hence the maximal distance any causal effects)
can propagate in one expansion time.

\begin{figure}
\begin{center}
\leavevmode
\epsfysize=9cm \epsfbox{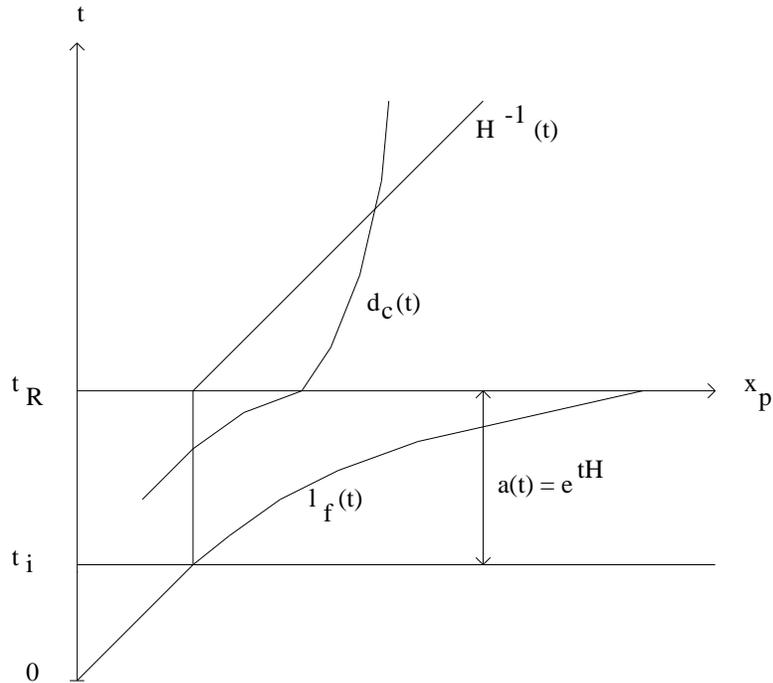}
\caption{
A sketch (physical coordinates vs. time) of the
solution of the formation of structure problem. Provided that the period of
inflation is sufficiently long, the separation $d_c$ between two galaxy
clusters is at all times smaller than the forward light cone. The dashed line
indicates the Hubble radius. Note that $d_c$ starts out smaller than the Hubble
radius, crosses it during the de Sitter period, and then reenters it at late
times.}
\end{center}
\end{figure}

As will be discussed in Section 4, the density perturbations produced during
inflation are due to quantum fluctuations in the matter and gravitational
fields \cite{Mukh80,Lukash}.  The amplitude of these inhomogeneities corresponds to a temperature $T_H$
\be
T_H \sim H \, , 
\ee
the Hawking temperature of the de Sitter phase. This leads one to expect that
at all times
during inflation, perturbations will be produced with a fixed physical wavelength $\sim
H^{-1}$. Subsequently, the length of the waves is stretched
with the expansion of space, and soon becomes larger than the Hubble radius.
The phases of the inhomogeneities are random.  Thus, the inflationary Universe
scenario predicts perturbations on all scales ranging from the comoving Hubble
radius at the beginning of inflation to the corresponding quantity at the time
of reheating.  In particular, provided that inflation lasts sufficiently long, perturbations on scales of galaxies and beyond will be generated. Note, however, that it is very dangerous to interpret de Sitter Hawking radiation as thermal radiation. In fact, the equation of state of this ``radiation" is not thermal \cite{RB83}.

\subsection{How to Obtain Inflation}

Obviously, the key question is how to obtain inflation. From the FRW equations, it follows that in order to get an exponential increase of the scale factor, the equation of state of matter must be
\be \label{infleos}
p = - \rho  
\ee
which is not compatible with the standard (cosmological) model description of matter as an ideal gas of classical matter.

As mentioned earlier, the ideal gas description of matter breaks down in the very early Universe. Matter must, instead, be described in terms of quantum field theory (QFT). In the resulting framework (classical general relativity as
a description of space and time, and QFT as a description of the matter content) it is possible to obtain inflation. More important than the quantum nature of matter is its {\it field} nature. Note, however, that quantum field-driven inflation is not the only way to obtain inflation. In fact, before the seminal paper by Guth \cite{Guth}, Starobinsky \cite{AS80} proposed a model with exponential expansion of the scale factor based on higher derivative curvature terms in the gravitational action.

Current quantum field theories of matter contain three types of fields: spin 1/2 fermions (the matter fields) $\psi$, spin 1 bosons $A_{\mu}$ (the gauge bosons) and spin 0 bosons, the scalar fields $\varphi$ (the Higgs fields used to spontaneously break internal gauge symmetries). The Lagrangian of the field theory is constrained by gauge invariance, minimal coupling and renormalizability. The Lagrangian of the bosonic sector of the theory is thus constrained to have the form
\be \label{fieldlag}
{\cal L}_m(\varphi, A_{\mu}) \, = \, {1 \over 2} D_{\mu}\varphi D^{\mu}\varphi - V(\varphi) + {1 \over 4} F_{\mu \nu}F^{\mu \nu} \, ,
\ee
where in Minkowski space-time $D_{\mu} = \partial_{\mu} - i g A_{\mu}$ denotes the (gauge) covariant derivative, $g$ being the gauge coupling constant, $F_{\mu \nu}$ is the field strength tensor, and $V(\varphi)$ is the Higgs potential. Renormalizability plus assuming symmetry under $\varphi \rightarrow - \varphi$ constrains $V(\varphi)$ to have the form
\be \label{pot}
V(\varphi) \, = \, {1 \over 2} m^2 \varphi^2 + {1 \over 4} \lambda \varphi^4 \, ,
\ee
where $m$ is the mass of the excitations of $\varphi$ about $\varphi = 0$, and $\lambda$ is a self-coupling constant. For spontaneous symmetry breaking,
$m^2 < 0$ is required.

Given the Lagrangian (\ref{fieldlag}), the action for matter is
\be
S_m \, = \int d^4x \sqrt{-g} {\cal L}_m \, ,
\ee
where $g$ here denotes the determinant of the metric tensor, and now the covariant derivative $D_{\mu}$ in (\ref{fieldlag}) is a gauge and metric covariant derivative. The energy-momentum tensor is obtained by varying this action with respect to the metric. The contributions of the scalar fields to the energy density $\rho$ and pressure $p$ are
\begin{eqnarray}
\rho (\varphi) & = & {1\over 2} \, \dot \varphi^2 + {1\over 2} \,
a^{-2}(\nabla \varphi)^2 + V (\varphi) \label{eos1} \\
p (\varphi) & = & {1\over 2} \dot \varphi^2 - {1\over 6} a^{-2}(\nabla \varphi)^2 - V(\varphi) \, . \label{eos2}
\end{eqnarray}
It thus follows that if the scalar field is homogeneous and static, but the potential energy positive, then the equation of state $p = - \rho$ necessary for exponential inflation results. This is the idea behind potential-driven inflation.

Note that given the restrictions imposed by minimal coupling, gauge invariance and renormalizability, scalar fields with nonvanishing potentials are required in order to obtain inflation. Mass terms for fermionic and gauge fields are not compatible with gauge invariance, and renormalizability forbids nontrivial potentials for fermionic fields. The initial hope of the inflationary Universe scenario \cite{Guth} was that the Higgs field required for gauge symmetry breaking in ``grand unified" (GUT) models would serve the role of the {\it inflaton}, the field generating inflation. As will be seen in the following subsection, this hope cannot be realized.

Most of the current realizations of potential-driven inflation are based on satisfying the conditions 
\be \label{srcond}
\dot \varphi^2, a^{-2} (\nabla \varphi)^2 \ll V (\varphi)\, , 
\ee
via the idea of slow rolling \cite{Linde82,AS82}. Consider the equation of motion of the scalar field $\varphi$ which can be obtained by varying the action $S_m$ with respect to $\varphi$:
\be \label{eom}
\ddot \varphi + 3 H \dot \varphi - a^{-2} \bigtriangledown^2 \varphi = -
V^\prime (\varphi)\, .  
\ee
If the scalar field starts out almost homogeneous and at rest, if the Hubble damping term (the second term on the l.h.s. of (\ref{eom}) is large), and if the potential is quite flat (so that the term on the r.h.s. of (\ref{eom}) is small), then ${\dot \varphi}^2$ may remain small compared to $V(\varphi)$, in which case the slow rolling conditions (\ref{srcond}) are satisfied and exponential inflation will result. If the spatial gradient terms are initially negligible, they will remain negligible since they redshift.

To illustrate the slow-roll inflationary scenario, consider the simplest model, a toy model with quadratic potential
\be
V(\varphi) \, = \, {1 \over 2} m^2 \varphi^2 \, .
\ee
Consider initial conditions for which $\varphi \gg m_{pl}$ and ${\dot \varphi} = 0$. At the beginning of the evolution, $\varphi$ will be rolling slowly and one finds approximate solutions by neglecting the $\ddot \varphi$ term (the self-consistency of this approximation needs to be checked in every model independently!).
In this simple model, the system of approximate equations 
\be
3 H {\dot \varphi} \, = \, - V^\prime (\varphi)\,
\ee
and  
\be
H^2 \, = \, {{8 \pi} \over 3} G V(\varphi)
\ee
can be solved exactly, yielding
\be \label{speed}
{\dot \varphi} = - {1 \over {\sqrt{12 \pi}}} m m_{pl} \, .
\ee
Since ${\dot \varphi}$ is constant, neglecting the ${\ddot \varphi}$ term in the equation of motion (\ref{eom}) is a self-consistent approximation. From (\ref{speed}) it also follows that inflation will occur until the slow-rolling condition (\ref{srcond}) breaks down, i.e. until
\be
\varphi \, = \, {1 \over {\sqrt{12 \pi}}} m_{pl} \, .
\ee
When $\varphi$ falls below the above value, it starts oscillating about its minimum with an amplitude that decays due to Hubble friction (the damping term $3 H {\dot \varphi}$ in the field equation of motion (\ref{eom})) and microscopic friction (see Section 4.1). Microscopic friction will lead to rapid heating of the Universe. For historic reasons, the time $t_R$ corresponding to the end of inflation and the onset of microscopic friction is called the {\it reheating time}.

The evolution of the scalar field $\varphi$ and of the temperature $T$ as a function of time is sketched in Figure 6.

\begin{figure}
\begin{center}
\leavevmode
\epsfysize=9cm \epsfbox{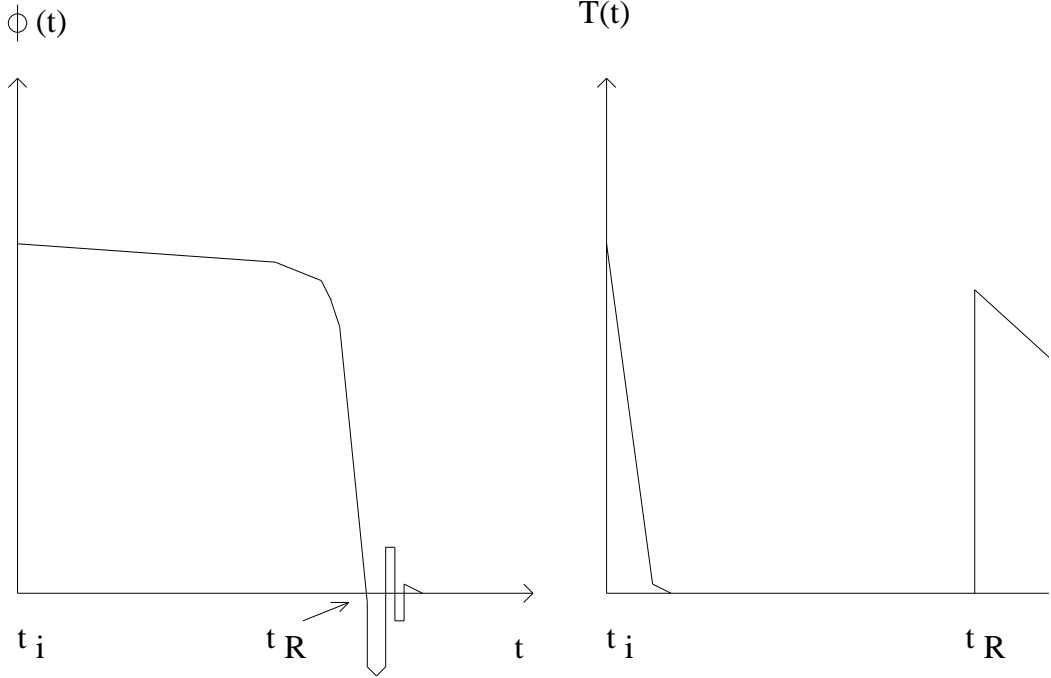}
\caption{ 
Evolution of $\varphi (t)$ and $T (t)$ in the
inflationary Universe scenario.}
\end{center}
\end{figure}
 
\subsection{Some Models of Inflation}

\medskip
\centerline{\bf Old Inflation}
\medskip

The first potential-driven model of inflation was, however, not based on slow rolling, but on false vacuum decay. It is
the ``Old Inflationary Universe" \cite{Guth,GuthTye} which was formulated in the context of a scalar field
theory which undergoes a first order phase transition.  As a toy
model, consider a scalar field theory with the potential $V (\varphi)$
of Figure 7.  This potential has a metastable ``false" vacuum at $\varphi = 0$, whereas the lowest energy state (the ``true" vacuum) is $\varphi = a$. Finite temperature effects \cite{finiteT} lead to extra terms in the effective potential which are proportional to $\varphi^2 T^2$ (the resulting finite temperature effective potential is also depicted in Figure 7). Thus, at high temperatures, the energetically preferred state is the false vacuum state. Note that this is only true if $\varphi$ is in thermal equilibrium with the other fields in the system.

The origin of the finite temperature corrections to the effective potential can be qualitatively understood as follows. Consider a theory with potential (\ref{pot}). If it is in thermal equilibrium, then the expectation value of  
$\varphi$ is given by
\be \label{expct}
<\varphi^2> \, \sim \, T^2 \, .
\ee
In the Hartree-Fock approximation, the interaction term $\lambda \varphi^3$ in the scalar field equation of motion (\ref{eom}) can be replaced by $3 \varphi <\varphi^2>$. Making use of (\ref{expct}) (with constant of proportionality designated by $\alpha$) to substitute for the expectation value of $\varphi^2$, we then get the same equation of motion as would follow for a scalar field with potential
\be
V_T(\varphi) \, = V(\varphi) + {3 \over 2} \alpha \lambda T^2 \phi^2 \, .
\ee
For a rigorous derivation, the reader is referred to the original articles \cite{finiteT} or the review article \cite{RB85}. However, from the heuristic analysis given above, it is already clear that the finite temperature corrections to the potential can only be applied if $\varphi$ is in thermal equilibrium.
 
\begin{figure}
\begin{center}
\leavevmode
\epsfxsize=8cm \epsfbox{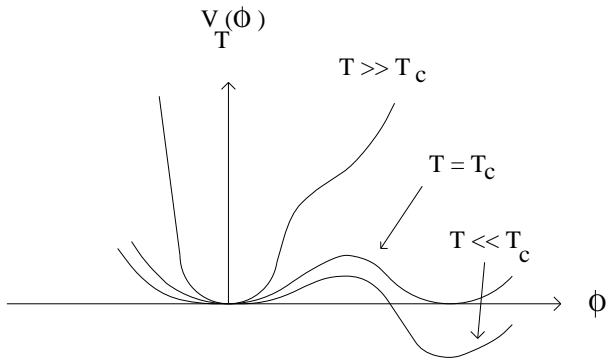}
\caption{
The finite temperature effective potential
in a theory with a first order phase transition.}
\end{center}
\end{figure}

For fairly general initial conditions, $\varphi (x)$ is trapped in the
metastable state $\varphi = 0$ as the Universe cools below the
critical temperature $T_c$.  As the Universe expands further, all
contributions to the energy-momentum tensor $T_{\mu \nu}$ redshift, except for
the contribution
\be
T_{\mu \nu} \sim V(\varphi) g_{\mu \nu} \, .
\ee
Hence, provided that the potential $V(\varphi)$ is shifted upwards such that $V(a) = 0$, the equation of state in the false vacuum approaches $p = - \rho$, and
inflation sets in. After a period $\Gamma^{-1}$, where $\Gamma$ is the tunnelling rate, bubbles of $\varphi = a$ begin to nucleate \cite{decay} in a sea of false
vacuum $\varphi = 0$. Inflation lasts until the false vacuum decays.
During inflation, the Hubble constant is given by
\be
H^2 = {8 \pi G\over 3} \, V (0) \, . 
\ee
The condition $V(a) = 0$, which looks rather unnatural, is required to
avoid a large cosmological constant today (none of the present inflationary Universe
models manage to circumvent or solve the cosmological constant problem).
 
It was immediately realized that old inflation has a serious ``graceful exit"
problem \cite{Guth,GuthWein}.  The bubbles nucleate after inflation with radius $r \ll 2t_R$. Even if the bubble walls expand with the speed of light, the bubbles would at the present time be much smaller than our apparent horizon.  Thus, unless bubbles percolate, the model predicts extremely large inhomogeneities inside
the Hubble radius, in contradiction with the observed isotropy of the
microwave background radiation.
\par
For bubbles to percolate, a sufficiently large number must be produced so that
they collide and homogenize over a scale larger than the present Hubble
radius.  However, because of the exponential expansion of the regions still in the false vacuum phase, the volume between bubbles
expands exponentially whereas the volume inside bubbles expands only with a low power. This prevents percolation. One way to overcome this problem is by realizing old inflation in the context of Brans-Dicke gravity \cite{MJ84,LS89}.

\medskip
\centerline{\bf New Inflation}
\medskip

Because of the graceful exit problem, old inflation never was considered to be
a viable cosmological model.  However, soon after the seminal paper by
Guth, Linde \cite{Linde82} and Albrecht and Steinhardt \cite{AS82}  independently put forwards a modified scenario, the ``New Inflationary Universe".

The starting point is a scalar field theory with a double well potential which
undergoes a second order phase transition (Fig. 8).  $V(\varphi)$ is
symmetric and $\varphi = 0$ is a local maximum of the zero temperature
potential.  Once again, it was argued that finite temperature effects confine
$\varphi(x)$ to values near $\varphi = 0$ at temperatures $T \geq
T_c$, where the critical temperature $T_c$ is characterized by the vanishing of the second derivative of $V_T(\varphi)$ at the origin.  For $T < T_c$, thermal fluctuations trigger the instability of $\varphi
(x) = 0$ and $\varphi (x)$ evolves towards either of the global minima at $\varphi = \pm \sigma$ by the classical equation of motion (\ref{eom}).
 
Within a fluctuation region, 
$\varphi(x)$ will be homogeneous. In such a region, we can  neglect the spatial gradient terms in Eq. (\ref{eom}).  Then, from (\ref{eos1}) and (\ref{eos2}) we can read off the
induced equation of state.  The slow rolling condition required to obtain inflation is given by (\ref{srcond}).

\begin{figure}
\begin{center}
\leavevmode
\epsfysize=7cm \epsfbox{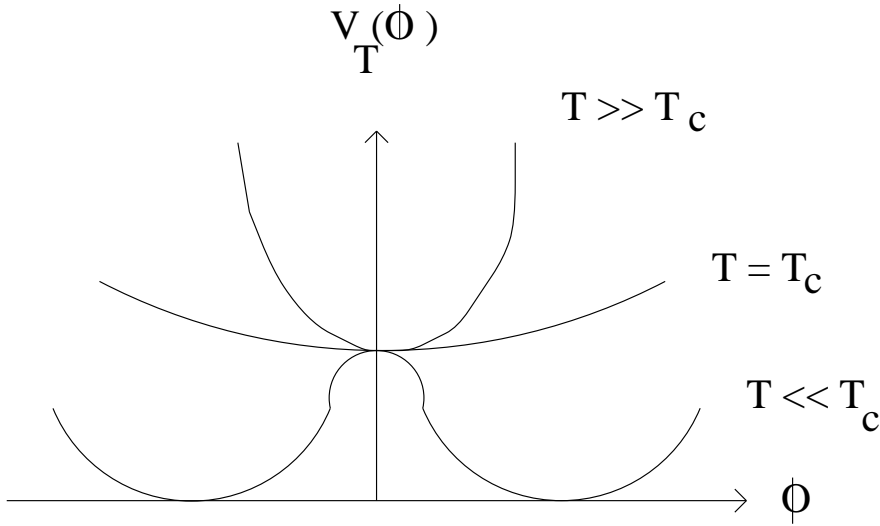}
\caption{ 
The finite temperature effective potential
in a theory with a second order phase transition.}
\end{center}
\end{figure}
 
There is no graceful exit problem in the new inflationary Universe.  Since the
fluctuation domains are established before the onset of inflation,
any boundary walls will be inflated outside the present Hubble radius.
 
In order to obtain inflation, the potential $V(\varphi)$ must be very flat near the false vacuum at $\varphi = 0$. This can only be the case if all of the coupling constants appearing in the potential are small. However, this implies that $\varphi$ cannot be in thermal equilibrium at early times, which would be required to localize $\varphi$ in the false vacuum. In the absence of thermal equilibrium, the initial conditions for $\varphi$ are only constrained by requiring that the total energy density in $\varphi$ not exceed the total energy density of the Universe. Most of the phase space of these initial conditions lies at values of $| \varphi | >> \sigma$. This leads to the ``chaotic" inflation scenario \cite{Linde83}.

\medskip  
\centerline{\bf Chaotic Inflation}
\medskip

Consider a region in space where at the initial time $\varphi (x)$
is very large, homogeneous and static.  In this case, the energy-momentum tensor will be
immediately dominated by the large potential energy term and induce an
equation of state $p \simeq - \rho$ which leads to inflation.  Due to the
large Hubble damping term in the scalar field equation of motion, $\varphi
(x)$ will only roll very slowly towards $\varphi = 0$.  The
kinetic energy contribution to $T_{\mu \nu}$ will remain small, the spatial
gradient contribution will be exponentially suppressed due to the expansion of
the Universe, and thus inflation persists. Note that in contrast to old and new inflation,
no initial thermal bath is required.  Note also that the precise form of
$V(\varphi)$ is irrelevant to the mechanism.  In particular, $V(\varphi)$ need
not be a double well potential.  This is a significant advantage, since for
scalar fields other than GUT Higgs fields used for spontaneous symmetry breaking,
there is no particle physics motivation for assuming a double well potential,
and the inflaton (the field which gives rise to inflation) cannot be a
conventional Higgs field, due to the severe fine tuning constraints.
 
The field and temperature evolution in a chaotic inflation model is similar to what is depicted in Figure 8, except that $\varphi$ is rolling towards the true vacuum at $\varphi = \sigma$ from the direction of large field values.

Chaotic inflation is a much more radical departure from standard cosmology than old and new inflation. In the latter, the inflationary phase can be viewed as a short phase of exponential expansion bounded at both ends by phases of radiation domination. In chaotic inflation, a piece of the Universe emerges with an inflationary equation of state immediately after the quantum gravity (or string) epoch.

The chaotic inflationary Universe scenario has been developed in great detail (see e.g. \cite{Linde94} for a recent review). One important addition is the inclusion of stochastic noise \cite{Starob87}  in the equation of motion for $\varphi$ in order to account for the effects of quantum fluctuations. In fact, it can be shown that for sufficiently large values of $|\varphi|$, the stochastic force terms are more important than the classical relaxation force $V^\prime(\varphi)$. There is thus equal probability for the quantum fluctuations to lead to an increase or decrease of $|\varphi|$. Hence, in a substantial fraction of the comoving volume, the field $\varphi$ climbs up the potential. This leads to the conclusion that chaotic inflation is eternal. At all times, a large fraction of the physical space will be inflating. Another consequence of including stochastic terms is that on large scales (much larger than the present Hubble radius), the Universe will look extremely inhomogeneous.

It is difficult to realize chaotic inflation in conventional supergravity models since gravitational corrections to the potential of scalar fields typically render the potential steep for values of $\vert \varphi \vert$ of the order of $m_{pl}$ and larger. This prevents the slow rolling condition (\ref{srcond}) from being realized. Even if this condition can be met, there are constraints from the amplitude of produced density fluctuations which are much harder to satisfy (see Section 5). Note that it is not impossible to obtain single field potential-driven inflation in supergravity models. For examples which show that is possible see e.g. \cite{ES95,ARS96}.

\medskip
\centerline{\bf Hybrid Inflation}
\medskip
Hybrid inflation \cite{hybrid} is a solution to the above-mentioned problem of chaotic inflation. Hybrid inflation requires at least two scalar fields to play an important role in the dynamics of the Universe. As a toy model, consider the potential of a theory with two scalar fields $\varphi$ and $\psi$:
\be
V(\varphi, \psi) \, = \, {1 \over 4} \lambda (M^2 - \psi^2)^2 + {1 \over 2} m^2 \varphi^2 + {1 \over 2} \lambda^{'} \psi^2 \varphi^2 \, .
\ee

For values of $\vert \varphi \vert$ larger than $\varphi_c$
\be
\varphi_c \, = \, \bigl({{\lambda} \over {\lambda^{'}}} M^2 \bigr)^{1/2} \, 
\ee
the minimum of $\psi$ is $\psi = 0$, whereas for smaller values of $\varphi$ the symmetry $\psi \rightarrow - \psi$ is broken and the ground state value of $\vert \psi \vert$ tends to $M$. The idea of hybrid inflation is that $\varphi$ is slowly rolling, like the inflaton field in chaotic inflation, but that the energy density of the Universe is dominated by $\psi$, i.e. by the contribution 
\be
V_0 \, = \, {1 \over 4} \lambda M^4 
\ee
to the potential. Inflation terminates once $\vert \varphi \vert$ drops below the critical value $\varphi_c$, at which point $\psi$ starts to move (and is not required to move slowly). 

Note that in hybrid inflation $\varphi_c$ can be much smaller than $m_{pl}$ and hence inflation without super-Planck scale values of the fields is possible. It is possible to implement hybrid inflation in the context of supergravity (see e.g. \cite{hybridSG}.
For a detailed discussion of inflation in the context of supersymmetric and supergravity models, the reader is referred to \cite{LR99}.

\medskip
\centerline{\bf Comments}
\medskip

At the present time there are many realizations of potential-driven inflation, but there is no canonical theory. A lot of attention is being devoted to implementing inflation in the context of unified theories, the prime candidate being superstring theory or M-theory. String theory or M-theory live in 10 or 11 space-time dimensions, respectively. When compactified to 4 space-time dimensions, there exist many {\it moduli} fields, scalar fields which describe flat directions in the complicated vacuum manifold of the theory. A lot of attention is now devoted to attempts at implementing inflation using moduli fields (see e.g. \cite{Banks} and references therein). 

Recently, it has been suggested that our space-time is a brane in a higher-dimensional space-time (see \cite{HW} for the basic construction).
Ways of obtaining inflation on the brane are also under active investigation (see e.g. \cite{braneinfl}).

It should also not be forgotten that inflation can arise from the purely gravitational sector of the theory, as in the original model of Starobinsky \cite{AS80} (see also Section 5), or that it may arise from kinetic terms in an effective action as in pre-big-bang cosmology \cite{PBB} or in k-inflation \cite{DM98}.  

\subsection{First Predictions of Inflation}

Theories with (almost) exponential inflation generically predict an (almost) scale-invariant spectrum of density fluctuations, as was first realized in \cite{Press,Mukh80,Lukash,Sato} and then studied more quantitatively in \cite{Mukh81,flucts,BST}. Via the Sachs-Wolfe effect \cite{SW}, these density perturbations induce CMB anisotropies with a spectrum which is also scale-invariant on large angular scales.

The heuristic picture is as follows (see Fig. 5). If the inflationary period which lasts from $t_i$ to $t_R$ is almost exponential, then the physical effects which are independent of the small deviations from exponential expansion are time-translation-invariant. This implies, for example, that quantum fluctuations at all times have the same strength when measured on the same physical length scale. 

If the inhomogeneities are small, they can described by linear theory, which implies that all Fourier modes $k$ evolve independently. The exponential expansion inflates the wavelength of any perturbation. Thus, the wavelength of perturbations generated early in the inflationary phase on length scales smaller than the Hubble radius soon becomes equal to (``exits") the  Hubble radius (this happens at the time $t_i(k)$) and continues to increase exponentially. After inflation, the Hubble radius increases as $t$ while the physical wavelength of a fluctuation increases only as $a(t)$. Thus, eventually the wavelength will cross the Hubble radius again (it will ``enter" the Hubble radius) at time $t_f(k)$. Thus, it is possible for inflation to generate fluctuations on cosmological scales by causal physics.

Any physical process which obeys the symmetry of the inflationary phase and which generates perturbations will generate fluctuations of equal strength when measured when they cross the Hubble radius:
\be
{{\delta M} \over M}(k, t_i(k)) \, = \, {\rm const}
\ee
(independent of $k$). Here, ${\delta M}(k, t)$ denotes the r.m.s. mass fluctuation on a length scale $k^{-1}$ at time $t$.

It is generally assumed that causal physics cannot affect the amplitude of fluctuations on super-Hubble scales (see, however, the comments at the end of Section 4.1). Therefore, the magnitude of ${{\delta M} \over M}$ can change only by a factor independent of $k$, and hence it follows that
\be \label{scaleinv}
{{\delta M} \over M}(k, t_f(k)) \, = \, {\rm const} \, ,
\ee
which is the definition of a scale-invariant spectrum \cite{HZ}. In terms of quantities usually used by astronomers, (\ref{scaleinv}) corresponds to a power spectrum
\be
P(k) \, \sim \, k \, .
\ee

Analyses from galaxy redshift surveys (see e.g. \cite{APM,LCRSPS}) give a power spectrum of density fluctuations which is consistent with a scale-invariant primordial spectrum as given by (\ref{scaleinv}). The COBE observations of CMB anisotropies \cite{COBE} are also in good agreement with the scale-invariant predictions from exponential inflation models, and in fact already give some bounds on possible deviations from scale-invariance. This agreement between the inflationary paradigm and observations is without doubt a major success of inflationary cosmology. However, it is worth pointing out (see \cite{VilShell,HK95,RB94} for recent reviews) that topological defect models also generically predict a scale-invariant spectrum of density fluctuations and CMB anisotropies. Luckily, the predictions of defect models and of inflationary theories differ in important ways: the small-scale CMB anisotropies are very different (see e.g. the lectures by Mageuijo in these proceedings \cite{MB99}), and the relative normalization of density and CMB fluctuations also differs. Observations will in the near future be able to discriminate between the predictions of inflationary cosmology and those of defect models.
   
\section{Progress in Inflationary Cosmology}

\subsection{Parametric Resonance and Reheating}

Reheating is an important stage in inflationary cosmology. It determines the state of the Universe after inflation and has consequences for baryogenesis, defect formation and other aspects of cosmology.

After slow rolling, the inflaton field begins to oscillate uniformly in space about the true vacuum state. Quantum mechanically, this corresponds to a coherent state of $k = 0$ inflaton particles. Due to interactions of the inflaton with itself and with other fields, the coherent state will decay into quanta of elementary particles. This corresponds to post-inflationary particle production.

Reheating is usually studied using simple scalar field toy models. The one we will adopt here consists of two real scalar fields, the inflaton $\varphi$
with Lagrangian
\be
{\cal L}_o \, = \, {1 \over 2} \partial_\mu \varphi \partial^\mu \varphi - {1 \over 4} \lambda (\varphi^2 - \sigma^2)^2 
\ee
interacting with a massless scalar field $\chi$ representing ordinary matter. The interaction Lagrangian is taken to be
\be
{\cal L}_I \, = \, {1 \over 2} g^2 \varphi^2 \chi^2 \, .
\ee
Self interactions of $\chi$ are neglected. 

By a change of variables
\be
\varphi \, = \, {\tilde \varphi} + \sigma \, ,
\ee
the interaction Lagrangian can be written as
\be \label{intlag}
{\cal L}_I \, = \, g^2 \sigma {\tilde \varphi} \chi^2 + {1 \over 2} g^2 {\tilde \varphi}^2 \chi^2 \, .
\ee
During the phase of coherent oscillations, the field ${\tilde \varphi}$ oscillates with a frequency
\be
\omega \, = \, m_{\varphi} \, = \, \lambda^{1/2} \sigma \, ,
\ee
neglecting the expansion of the Universe, although this can be taken into account \cite{KLS94,STB95}).

In the {\it elementary theory of reheating} (see e.g. \cite{DolLin} and \cite{AFW}), the decay of the inflaton is calculated using first order perturbation theory. According to the Feynman rules, the decay rate $\Gamma_B$ of $\varphi$ (calculated assuming that the cubic coupling term dominates) is
given by
\be
\Gamma_B \, = \, {{g^2 \sigma^2} \over {8 \pi m_{\phi}}} \, .
\ee
The decay leads to a decrease in the amplitude of $\varphi$ (from now on we will drop the tilde sign) which can be approximated by adding an extra damping term to the equation of motion for $\varphi$:
\be
{\ddot \varphi} + 3 H {\dot \varphi} + \Gamma_B {\dot \varphi} \, = \,
- V^\prime(\varphi) \, .
\ee
From the above equation it follows that as long as $H > \Gamma_B$, particle production is negligible. During the phase of coherent oscillation of $\varphi$, the energy density and hence $H$ are decreasing. Thus, eventually $H = \Gamma_B$, and at that point reheating occurs (the remaining energy density in $\varphi$ is very quickly transferred to $\chi$ particles.

The temperature $T_R$ at the completion of reheating can be estimated by computing the temperature of radiation corresponding to the value of $H$ at which $H = \Gamma_B$. From the FRW equations it follows that
\be
T_R \, \sim \, (\Gamma_B m_{pl})^{1/2} \, .
\ee
If we now use the ``naturalness" constraint{\footnote{At one loop order, the cubic interaction term will contribute to $\lambda$ by an amount $\Delta \lambda \sim g^2$. A renormalized value of $\lambda$ smaller than $g^2$ needs to be finely tuned at each order in perturbation theory, which is ``unnatural".}} 
\be
g^2 \, \sim \, \lambda
\ee
in conjunction with the constraint on the value of $\lambda$ from (\ref{lambdaconstr}), it follows that for $\sigma < m_{pl}$,
\be
T_R \, < \, 10^{10} {\rm GeV} \, .
\ee
This would imply no GUT baryogenesis, no GUT-scale defect production, and no gravitino problems in supersymmetric models with $m_{3/2} > T_R$, where $m_{3/2}$ is the gravitino mass. As we shall see, these conclusions change radically if we adopt an improved analysis of reheating.

As was first realized in \cite{TB90}, the above analysis misses an essential point. To see this, we focus on the equation of motion for the matter field $\chi$ coupled to the inflaton $\varphi$ via the interaction Lagrangian ${\cal L}_I$ of (\ref{intlag}). Considering only the cubic interaction term, the equation of motion becomes
\be
{\ddot \chi} + 3H{\dot \chi} - \bigl(({{\nabla} \over a})^2 - m_{\chi}^2 - 2g^2\sigma\varphi \bigr)\chi \, = \, 0 \, .
\ee
Since the equation is linear in $\chi$, the equations for the Fourier modes $\chi_k$ decouple:
\be \label{reseq}
{\ddot \chi_k} + 3H{\dot \chi_k} + (k_p^2 + m_{\chi}^2 + 2g^2\sigma\varphi)\chi_k \, = \, 0 ,
\ee
where $k_p = k / a$ is the time-dependent physical wavenumber. 

Let us for the moment neglect the expansion of the Universe. In this case, the friction term in (\ref{reseq}) drops out and $k_p$ is time-independent, and Equation (\ref{reseq}) becomes a harmonic oscillator equation with a time-dependent mass determined by the dynamics of $\varphi$. In the reheating phase, $\varphi$ is undergoing oscillations. Thus, the mass in (\ref{reseq}) is varying periodically. In the mathematics literature, this equation is called the Mathieu equation. It is well known that there is an instability. In physics, the effect is known as {\bf parametric resonance} (see e.g. \cite{parres}). At frequencies $\omega_n$ corresponding to half integer multiples of the frequency $\omega$ of
the variation of the mass, i.e.
\be
\omega_k^2 = k_p^2 + m_{\chi}^2 \, = \, ({n \over 2} \omega)^2 \,\,\,\,\,\,\, n = 1, 2, ... ,
\ee
there are instability bands with widths $\Delta \omega_n$. For values of $\omega_k$ within the instability band, the value of $\chi_k$ increases exponentially:
\be
\chi_k \, \sim \, e^{\mu t} \,\,\,\, {\rm with} \,\,\, \mu \sim {{g^2 \sigma \varphi_0} \over {\omega}} \, ,
\ee
with $\varphi_0$ being the amplitude of the oscillation of $\varphi$. Since the widths of the instability bands decrease as a power of the (small) coupling constant $g^2$ with increasing $n$, for practical purposes only the lowest instability band is important. Its width is
\be
\Delta \omega_k \, \sim \, g \sigma^{1/2} \varphi_0^{1/2} \, .
\ee
Note, in particular, that there is no ultraviolet divergence in computing the total energy transfer from the $\varphi$ to the $\chi$ field due to parametric resonance.

It is easy to include the effects of the expansion of the Universe (see e.g. \cite{TB90,KLS94,STB95}). The main effect is that the value of $\omega_k$ can become time-dependent. Thus, a mode may enter and leave the resonance bands. In this case, any mode will lie in a resonance band for only a finite time. This can reduce the efficiency of parametric resonance, but the amount of reduction is quite dependent on the specific model. This behavior of the modes, however, also has positive aspects: it implies \cite{TB90} that the calculation of energy transfer is perfectly well-behaved and no infinite time divergences arise.

It is now possible to estimate the rate of energy transfer, whose order of magnitude is given by the phase space volume of the lowest instability band multiplied by the rate of growth of the mode function $\chi_k$. Using as an initial condition for $\chi_k$ the value $\chi_k \sim H$ given by the magnitude of the expected quantum fluctuations, we obtain
\be \label{entransf}
{\dot \rho} \, \sim \, \mu ({\omega \over 2})^2 \Delta\omega_k H e^{\mu t} \, .
\ee

From (\ref{entransf}) it follows that provided that the condition
\be \label{rescond}
\mu \Delta t \, >> 1
\ee
is satisfied, where $\Delta t < H^{-1}$ is the time a mode spends in the instability band, then the energy transfer will procede fast on the time scale
of the expansion of the Universe. In this case, there will be explosive particle production, and the energy density in matter at the end of reheating will be approximately equal to the energy density at the end of inflation.  

The above is a summary of the main physics of the modern theory of reheating.
The actual analysis can be refined in many ways (see e.g. \cite{KLS94,STB95,KLS97}).
First of all, it is easy to take the expansion of the Universe into account
explicitly (by means of a transformation of variables), to employ an exact solution of the background model and to reduce the mode equation for $\chi_k$ to a Hill equation, an equation similar to the Mathieu equation which also admits exponential instabilities.

The next improvement consists of treating the $\chi$ field quantum mechanically (keeping $\varphi$ as a classical background field). At this point, the techniques of quantum field theory in a curved background can be applied. There is no need to impose artificial classical initial conditions for $\chi_k$. Instead, we may assume that $\chi$ starts in its initial vacuum state (excitation of an initial thermal state has been studied in \cite{Yoshimura2}), and the Bogoliubov mode mixing technique (see e.g. \cite{Birrell}) can be used to compute the number of particles at late times.

Using this improved analysis, we recover the result (\ref{entransf}) \cite{STB95}. Thus, provided that the condition (\ref{rescond}) is satisfied, reheating will be explosive. Working out the time $\Delta t$ that a mode remains in the instability band for our model, expressing $H$ in terms of $\varphi_0$ and $m_{pl}$, and $\omega$ in terms of $\sigma$, and using the naturalness relation $g^2 \sim \lambda$, the condition for explosive particle production becomes
\be \label{rescond2}
{{\varphi_0 m_{pl}} \over {\sigma^2}} \, >> \, 1 \, ,
\ee
which is satisfied for all chaotic inflation models with $\sigma < m_{pl}$ (recall that slow rolling ends when $\varphi \sim m_{pl}$ and that therefore the initial amplitude $\varphi_0$ of oscillation is of the order $m_{pl}$).

We conclude that rather generically, reheating in chaotic inflation models will be explosive. This implies that the energy density after reheating will be approximately equal to the energy density at the end of the slow rolling period. Therefore, as suggested in \cite{KLS96,Tkachev} and \cite{KLR96}, respectively, GUT scale defects may be produced after reheating and GUT-scale baryogenesis scenarios may be realized, provided that the GUT energy scale is lower than
the energy scale at the end of slow rolling.

Note that the state of $\chi$ after parametric resonance is {\bf not} a thermal state. The spectrum consists of high peaks in distinct wave bands. An important question which remains to be studied is how this state thermalizes.
For some interesting work on this issue see \cite{therm}. As emphasized in \cite{KLS96} and \cite{Tkachev}, the large peaks in the spectrum may lead to symmetry restoration and to the efficient production of topological defects (for a differing view on this issue see \cite{KK97,PS98}). Since the state after explosive particle production is not a thermal state, it is useful to follow
\cite{KLS94} and call this process ``preheating" instead of reheating.

Note that the details of the analysis of preheating are quite model-dependent. In fact$^{\cite{KLS94,KLS97}}$, in most models one does not get the kind of ``narrow-band" resonance discussed here, but ``wide-band" resonance. In this case, the energy transfer is even more efficient.

Many important questions, e.g. concerning thermalization and back-reaction effects during and after preheating (or parametric resonance) remain to be fully analyzed. Recently \cite{BKM1} it has been argued that parametric resonance may lead to resonant amplification of super-Hubble-scale cosmological perturbations and might possibly even modify some of the first predictions of inflation mentioned in Section 3.4. The point is that in the presence of an oscillating inflaton field, the equation of motion for the cosmological perturbations takes on a similar form to the Mathieu equation discussed above (\ref{reseq}). In some models of inflation, the first resonance band included modes with wavelength larger than the Hubble radius, leading to the apparent amplification of super-Hubble-scale modes which will destroy the scale-invariance of the fluctuations. Such a process would not violate causality \cite{FB99} since it is driven by the inflaton field which is coherent on super-Hubble scales at the end of inflation as a consequence of the causal dynamics of an inflationary Universe. However, careful analyses for simple single-field \cite{FB99,PE99} and double-field \cite{JS99,PI99} models demonstrated that there is no net growth of the physical amplitude of gravitational fluctuations beyond what the usual theory of cosmological perturbations (see the following subsection) predicts. It is still possible, however, that in more complicated model a net physical effect of parametric resonance of gravitational fluctuations persists \cite{BKM2}.

\subsection{Quantum Theory of Cosmological Perturbations}

On scales larger than the Hubble radius $(\lambda > t)$ the Newtonian theory of
cosmological perturbations obviously is inapplicable, and a general
relativistic analysis is needed.  On these scales, matter is essentially frozen
in comoving coordinates.  However, space-time fluctuations can still increase
in amplitude.

In principle, it is straightforward to work out the general relativistic theory
of linear fluctuations \cite{Lifshitz}.  We linearize the Einstein  equations
\be
G_{\mu\nu} = 8 \pi G T_{\mu\nu} 
\ee
(where $G_{\mu\nu}$ is the Einstein tensor associated with the space-time
metric $g_{\mu\nu}$, and $T_{\mu\nu}$ is the energy-momentum tensor of matter)
about an expanding FRW background $(g^{(0)}_{\mu\nu} ,\, \varphi^{(0)})$:
\begin{eqnarray}
g_{\mu\nu} (\underline{x}, t) & = & g^{(0)}_{\mu\nu} (t) + h_{\mu\nu}
(\underline{x}, t) \\
\varphi (\underline{x}, t) & = & \varphi^{(0)} (t) + \delta \varphi
(\underline{x}, t) \,  
\end{eqnarray}
and pick out the terms linear in $h_{\mu\nu}$ and $\delta \varphi$ to obtain
\be \label{linein}
\delta G_{\mu\nu} \> = \> 8 \pi G \delta T_{\mu\nu} \, . 
\ee
In the above, $h_{\mu\nu}$ is the perturbation in the metric and $\delta
\varphi$ is the fluctuation of the matter field $\varphi$.  We have denoted all
matter fields collectively by $\varphi$.

In practice, there are many complications which make this analysis highly
nontrivial.  The first problem is ``gauge invariance" \cite{PressVish}   Imagine starting
with a homogeneous FRW cosmology and introducing new coordinates which mix
$\underline{x}$ and $t$.  In terms of the new coordinates, the metric now looks
inhomogeneous.  The inhomogeneous piece of the metric, however, must be a pure
coordinate (or "gauge") artefact.  Thus, when analyzing relativistic
perturbations, care must be taken to factor out effects due to coordinate
transformations.

There are various methods of dealing with gauge artefacts.  The simplest and
most physical approach is to focus on gauge invariant variables, i.e.,
combinations of the metric and matter perturbations which are invariant under
linear coordinate transformations.

The gauge invariant theory of cosmological perturbations is in principle
straightforward, although technically rather tedious. In the following I will
summarize the main steps and refer the reader to \cite{MFB92} for the details and further references (see also \cite{MFB92} for a pedagogical introduction and \cite{Bardeen,BKP83,KoSa84,Durrer,Lyth,Hwang,EllisBruni,Salopek} for other approaches).

We consider perturbations about a spatially flat Friedmann-Robertson-Walker
metric
\be
ds^2 = a^2 (\eta) (d\eta^2 - d \underline{x}^2) 
\ee
where $\eta$ is conformal time (related to cosmic time $t$ by $a(\eta)  d \eta
= dt$).  At the linear level, metric perturbations can be decomposed into scalar modes, vector modes and tensor modes (gravitational waves). In the following, we will focus on the scalar modes since they are the only ones which couple to energy density and pressure. A scalar metric perturbation (see \cite{Stewart} for a precise definition)
can be written in terms of four free functions of space and time:
\be
\delta g_{\mu\nu} = a^2 (\eta) \pmatrix{2 \phi & -B_{,i} \cr
-B_{,i} & 2 (\psi \delta_{ij} + E_{,ij}) \cr} \, . 
\ee
 
The next step is to consider infinitesimal coordinate transformations
which preserve the scalar nature of $\delta g_{\mu\nu}$, and to calculate the
induced transformations of $\phi, \psi, B$ and $E$.  Then we find invariant
combinations to linear order.  (Note that there are in general no combinations
which are invariant to all orders \cite{SteWa}.)  After some algebra, it follows
that
\begin{eqnarray}
\Phi & = & \phi + a^{-1} [(B - E^\prime) a]^\prime \\
\Psi & = & \psi - {a^\prime\over a} \, (B - E^\prime)  
\end{eqnarray}
are two invariant combinations (a prime denotes differentiation
with respect to $\eta$).

Perhaps the simplest way \cite{MFB92} to derive the equations of motion for gauge invariant variables is to consider the linearized
Einstein equations (\ref{linein}) and to write them out in the longitudinal gauge defined by $B = E = 0$, in which $\Phi = \phi$ and $\Psi = \psi$, to directly obtain gauge invariant equations.

For several types of matter, in particular for scalar field matter,  
$\delta T^i_j \sim \delta^i_j$ 
which implies $\Phi = \Psi$.  Hence, in this case the scalar-type cosmological perturbations can be described by a single gauge invariant variable.  The
equation of motion takes the form \cite{BST,BK84,Lyth,RBrev,Gotz} 
\be \label{conserv}
\dot \xi = O \left({k\over{aH}} \right)^2 H \xi  
\ee
where
\be
\xi = {2\over 3} \, {H^{-1} \dot \Phi + \Phi\over{1 + w}} + \Phi \, . 
\ee

The variable $w = p/ \rho$ (with $p$ and $\rho$ background pressure and energy
density respectively) is a measure of the background equation of state.  In
particular, on scales larger than the Hubble radius, the right hand side of
(\ref{conserv}) is negligible, and hence $\xi$ is constant.

If the equation of state of matter is constant, {\it i.e.}, $w = {\rm
const}$, then $\dot \xi = 0$ implies that the relativistic potential
is time-independent on scales larger than the Hubble radius, i.e. $\Phi (t) = {\rm const}$. During a transition from an
initial phase with $w = w_i$  to a phase with $w = w_f$, $\Phi$ changes. In many cases, a good approximation to the dynamics given by (\ref{conserv}) is
\be \label{cons3}
{\Phi\over{1 + w}}(t_i)  \, = \, {\Phi\over{1 + w}}(t_f)  \, , 
\ee

In order to make contact with matter perturbations and Newtonian intuition, it
is important to remark that,  as a consequence of the Einstein constraint
equations, at Hubble radius crossing $\Phi$ is a measure of the fractional
density fluctuations:
\be
\Phi (k, t_H (k) ) \sim {\delta \rho\over \rho} \, ( k , \, t_H (k) ) \, .
\ee
 
As mentioned earlier, the primordial fluctuations in an inflationary cosmology are generated by quantum fluctuations. What follows is a very brief description of the unified
analysis of the quantum generation and evolution of perturbations in an inflationary Universe (for a detailed review see \cite{MFB92}).
The basic point is that at the linearized level, the equations describing both gravitational and matter perturbations can be quantized in a consistent way. The use of gauge invariant variables makes the analysis both physically clear and computationally simple. 

The first step of this analysis is to consider the action for the linear perturbations in a background homogeneous and isotropic Universe, i.e. to expand the gravitational and matter action $S(g_{\mu \nu}, \varphi)$ to quadratic order in the fluctuation variables $h_{\mu \nu}, \delta \varphi$
\be
S(g_{\mu \nu}, \varphi) \, = \, S_0(g^{(0)}_{\mu\nu}, \varphi^{(0)}) \, + S_2(h_{\mu \nu}, \delta \varphi; g^{(0)}_{\mu\nu}, \varphi^{(0)}) \, ,
\ee
where $S_2$ is quadratic in the perturbation variables. Focusing on the scalar perturbations, it turns out that one can express the resulting $S_2$ in terms of the joint metric and matter gauge invariant variable
\be
v \, = \, a \bigl( \delta \varphi + {{\varphi^{(0), \prime}} \over {\cal{H}}} \Phi \bigr)
\ee
describing the fluctuations. In the above, a prime denotes the derivative with respect to conformal time, and ${\cal H} \, = \, a^{\prime} / a$. It turns out that, after a lot of algebra, the action $S_2$ reduces to the action of a single gauge invariant free scalar field (namely $v$) with a time dependent mass \cite{Mukh88,Sasaki}  (the time dependence reflects the expansion of the background space-time) 
\be
S_2 \, = \, {1 \over 2} \int dt d^3x \bigl( v^{\prime 2} - (\nabla v)^2 + {{z^{\prime \prime}} \over z} v^2 \bigr) \, ,
\ee
with
\be
z \, = \, {{a \varphi_0^{\prime}} \over {\cal H}} \, .
\ee
This result is not surprising. Based on the study of classical cosmological perturbations, we know that there is only one field degree of freedom for the scalar perturbations. Since at the linearized level there are no mode interactions, the action for this field must be that of a free scalar field. 

The action thus has the same form as the action for a free scalar matter field in a time dependent gravitational or electromagnetic background, and we can use standard methods to quantize this theory (see e.g. \cite{Birrell}). If we employ canonical quantization, then the mode functions of the field operator obey the same classical equations as we derived in the gauge-invariant analysis of relativistic perturbations. 

The time dependence of the mass is reflected in the nontrivial form of the solutions of the mode equations. The mode equations have growing modes which correspond to particle production or equivalently to the generation and amplification of fluctuations. We can start the system off (e.g. at the beginning of inflation) in the vacuum state (defined as a state with no particles with respect to a local comoving observer). The state defined this way will not be the vacuum state from the point of view of an observer at a later time. The Bogoliubov mode mixing technique (see e.g. \cite{Birrell} for a detailed exposition) can be used to calculate the number density of particles at a later time. In particular, expectation values of field operators such as the power spectrum can be computed.

The resulting power spectrum gives the following result for the mass perturbations at time $t_i(k)$:
\be \label{inmass}
\left( {\delta M\over M} \right)^2 \, (k, t_i (k)) \sim k^3 \left({V^\prime
(\varphi_0) \delta \tilde \varphi (k, t_i (k))\over \rho_0} \right)^2 \sim
\left({V^\prime (\varphi_0) H\over \rho_0 } \right)^2 \, . 
\ee
 
If the background scalar field is rolling slowly, then
\be \label{slowroll1}
V^\prime (\varphi_0 (t_i (k))) =  3 H | \dot \varphi_0 (t_i (k)) | \, .
\ee
and
\be \label{slowroll2}
(1 + p/\rho)(t_i(k)) \, \simeq \, \rho_0^{-1} {\dot \varphi_0^2}(t_i(k)) \, .
\ee
Combining (\ref{cons3}), (\ref{inmass}), (\ref{slowroll1}) and (\ref{slowroll2}) and we get
\be
{\delta M\over M} (k, \, t_f (k))  \sim \, {3 H^2 | \dot \varphi_0
(t_i (k)) |\over{\dot \varphi^2_0 (t_i (k))}} =  {3H^2\over{| \dot \varphi_0 (t_i (k))|}} 
\ee
This result can now be evaluated for specific models of inflation to find the
conditions on the particle physics parameters which give a value
\be \label{obs2}
{\delta M\over M} (k, \, t_f (k))  \sim 10^{-5} 
\ee
which is required if quantum fluctuations from inflation are to provide the
seeds for galaxy formation and agree with the CMB anisotropy limits.

For chaotic inflation with a potential
\be
V (\varphi) = {1\over 2} m^2 \varphi^2 \, , 
\ee
we can solve the slow rolling equations for the inflaton to obtain
\be \label{massconstr}
{\delta M\over M} (k, t_f (k))  \sim 10^2 {m\over m_{pl}} 
\ee
which implies that $m \sim 10^{13} \, {\rm GeV}$ to agree with (\ref{obs2}).

Similarly, for a quartic potential  
\be \label{pot2}
V (\varphi) = {1\over 4} \lambda \varphi^4 
\ee
we obtain
\be \label{lambdaconstr}
{\delta M\over M} (k, \, t_f (k)) \sim  10^2 \cdot \lambda^{1/2} 
\ee
which requires $\lambda \leq 10^{-12}$ in order not to conflict with observations.

Demanding that (\ref{massconstr}) and (\ref{lambdaconstr}) yield the observed amplitude of the density perturbations requires the presence of small parameters in the particle physics models.  
It  has been shown \cite{Freese} that, quite generally,  
small parameters are required in any particle physics model if potential-driven inflation is to solve the fluctuation problem.

To summarize the main results of the analysis of density
fluctuations in inflationary cosmology:
\begin{enumerate}
\item{} Quantum vacuum fluctuations in the de Sitter phase of an inflationary
Universe are the source of perturbations.
\item{} As a consequence of the change in the background equation of state, the evolution outside the Hubble radius produces a large
amplification of the perturbations.  In fact, unless the particle physics
model contains very small coupling constants, the predicted fluctuations are
in excess of those allowed by the bounds on cosmic microwave anisotropies.
\item{} The quantum generation and classical evolution of fluctuations can be treated in a unified manner. The formalism is no more complicated that the study of a free scalar field in a time dependent background.
\item{} Inflationary Universe models generically produce an approximately scale invariant Harrison-Zel'dovich spectrum
\be
{\delta M\over M} (k , t_f (k) ) \, \simeq \, {\rm const.} 
\ee
\end{enumerate}

It is not hard to construct models which give a different spectrum.  All that
is required is a significant change in $H$ during the period of
inflation. 

\section{Problems of Inflationary Cosmology}

\subsection{Fluctuation Problem}

A generic problem for all realizations of potential-driven inflation studied up to now concerns the amplitude of the density perturbations which are induced by quantum fluctuations during the period of exponential expansion \cite{flucts,BST}. From the amplitude of CMB anisotropies measured by COBE, and from the present amplitude of density inhomogeneities on length scales of clusters of galaxies, it follows that the amplitude of the mass fluctuations ${\delta M} / M$ on a length scale given by the comoving wavenumber $k$ at the time $t_H(k)$ when that scale crosses the Hubble radius in the FRW period is of the order $10^{-4}$. 

However, as was discussed in detail in the previous section, the present realizations of inflation based on scalar quantum field matter generically \cite{Freese} predict a much larger value of these fluctuations, unless a parameter in the scalar field potential takes on a very small value. For example, in a single field chaotic inflationary model with potential
given by (\ref{pot2}) the mass fluctuations generated are of the order
$10^2 \lambda^{1/2}$ (see (\ref{lambdaconstr})). Thus, in order not to conflict with observations, a value of $\lambda$ smaller than $10^{-12}$ is required. There have been many attempts to justify such small parameters based on specific particle physics models, but no single convincing model has emerged.

\subsection{Super-Planck-Scale Physics Problem}

In many models of inflation, in particular in chaotic inflation, the period of inflation is so long that comoving scales of cosmological interest today corresponded to a physical wavelength much smaller than the Planck length at the beginning of inflation. In extrapolating the evolution of cosmological perturbations according to linear theory to these very early times, we are implicitly making the assumption that the theory remains perturbative to arbitrarily high energies. If there were completely new physics at the Planck scale, the predictions might change. For example, if there were a sharp ultraviolet cutoff in the theory, then, if inflation lasts many e-folding, the modes which represent fluctuations on galactic scales today would not be present in the theory since their wavelength would have been smaller than the cutoff length at the beginning of inflation.  A similar concern about black hole Hawking radiation has been raised in \cite{Jacobson}.

As an example of how Planck-scale physics may dramatically alter the usual predictions of inflation, consider ``Pre-big-bang Cosmology" \cite{PBB} which can be viewed as a toy model for how to
include some effects of string theory in cosmological considerations. The pre-big-bang scenario is based on a dilaton-dominated super-exponentially expanding Universe smoothly connecting to an expanding FRW Universe dominated by matter and radiation. In this model of the early Universe, scalar metric perturbations on large scales are highly suppressed \cite{PBBflucts} in the absence of excited axionic degrees of freedom \cite{PBBaxion}.

\subsection{Singularity Problem}

Scalar field-driven inflation does not eliminate singularities from cosmology. Although the standard assumptions of the Penrose-Hawking theorems break down if matter has an equation of state with negative pressure, as is the case during inflation, nevertheless it can be shown that an initial singularity persists in inflationary cosmology \cite{Borde}. This implies that the theory is incomplete. In particular, the physical initial value problem is not defined.
 
\subsection{Cosmological Constant Problem}

Since the cosmological constant acts as an effective energy density, its value is bounded from above by the present energy density of the Universe. In Planck units, the constraint on the effective cosmological constant $\Lambda_{eff}$ is
(see e.g. \cite{cosmorev})
\be
{{\Lambda_{eff}} \over {m_{pl}^4}} \, \le \, 10^{- 122} \, .
\ee
This constraint applies both to the bare cosmological constant and to any matter contribution which acts as an effective cosmological constant.

The true vacuum value of the potential $V(\varphi)$ acts as an effective cosmological constant. Its value is not constrained by any particle physics requirements (in the absence of special symmetries). The cosmological constant problem is thus even more acute in inflationary cosmology than it usually is. The same unknown mechanism which must act to shift the potential such that inflation occurs in the false vacuum must also adjust the potential to vanish in the true vacuum. 
Supersymmetric theories may provide a resolution of this problem, since unbroken supersymmetry forces $V(\varphi) = 0$ in the supersymmetric vacuum. However, supersymmetry breaking will induce a nonvanishing $V(\varphi)$ in the true vacuum after supersymmetry breaking.

\section{New Avenues}

In the light of the problems of potential-driven inflation discussed in the previous sections, many cosmologists have begun thinking about new avenues towards early Universe cosmology which, while maintaining (some of) the successes of inflation, address and resolve some of its difficulties. One approach which has received a lot of recent attention is pre-big-bang cosmology \cite{PBB}. A nice feature of this theory is that the mechanism of inflation is completely independent of a potential and thus independent of the cosmological constant issue. The scenario, however, is confronted with a graceful exit problem \cite{PBBEP}, and the initial conditions need to be very special \cite{PBBIC} (see, however, the
discussion in \cite{BDV}). String theory may lead to a natural resolution of some of the puzzles of inflationary cosmology. This is an area of active research. The reader is referred to \cite{Banks} for a review of recent studies of obtaining inflation with moduli fields, and to \cite{braneinfl} for attempts to obtain inflation with branes. Below, three more conventional approaches to addressing some of the problems of inflation will be summarized.
 
\subsection{Inflation from Condensates}

At the present time there is no direct observational evidence for the existence of fundamental scalar fields in nature (in spite of the fact that most attractive unified theories of nature require the existence of scalar fields in the low energy effective Lagrangian). Scalar fields were initially introduced in particle physics to yield an order parameter for the symmetry breaking phase transition. Many phase transitions exist in nature; however, in all cases, the order parameter is a condensate. Hence, it is useful to consider the possibility of obtaining inflation using condensates, and in particular to ask if this would yield a different inflationary scenario.

The analysis of a theory with condensates is intrinsically non-perturbative. The expectation value of the Hamiltonian $\la H \ra$ of the theory contains terms with arbitrarily high powers of the expectation value $\la \varphi \ra$ of the condensate. A recent study of the possibility of obtaining inflation in a theory with condensates was undertaken in \cite{BZ97} (see also \cite{BM92} for some earlier work). Instead of truncating the expansion of $\la H \ra$ at some arbitrary order, the assumption was made that the expansion of $\la H \ra$ in powers of $\la \varphi \ra$ is asymptotic and, specifically, Borel summable (on general grounds one expects that the expansion will be asymptotic - see e.g. \cite{ARZ})
\begin{eqnarray} \label{BZpot}
\la H\ra \, &=& \, \sum_{n=0}^{\infty} {n!} (-1)^n a_n \la \varphi^n \ra \nonumber \\
&=& \, \int_0^{\infty} ds {{f(s)} \over {s(sm_{pl} + \la \varphi \ra)}} e^{-1/s} \, .
\end{eqnarray}

The cosmological scenario is as follows: the expectation value $\la \varphi \ra$ vanishes at times before the phase transition when the condensate forms. Afterwards, $\la \varphi \ra$ evolves according to the classical equations of motion with the potential given by (\ref{BZpot}) (we have no information about the form of the kinetic term but will assume that it takes the standard form).
Hence, the initial conditions for the evolution of $\la \varphi \ra$ are like those of new inflation. It can be easily checked that the slow rolling conditions are satisfied. However, the slow roll conditions remain satisfied for all values of $\la \varphi \ra$, thus leading to a graceful exit problem - inflation will never terminate.

However, we have neglected the fact that correlation functions, in particular $\la \phi^2 \ra$, are in general infrared divergent in the de Sitter phase of an expanding Universe. It is natural to introduce a phenomenological cutoff parameter $\epsilon(t)$ into the vacuum expectation value (VEV), and to replace $\la \varphi \ra$ by $\la \varphi \ra \, / \, \epsilon$. It is natural to expect that $\epsilon(t) \sim H(t)$ (see e.g. \cite{AL82b,VilFord}). Hence, the dynamical system consists of two coupled functions of time $\la \varphi \ra$ and $\epsilon$. A careful analysis shows that a graceful exit from inflation occurs precisely if $\la H \ra$ tends to zero when $\la \varphi \ra$ tends to large values. 

As is evident, the scenario for inflation in this composite field model is very different from the standard potential-driven inflationary scenario. It is particularly interesting that the graceful exit problem from inflation is linked to the cosmological constant problem.
 
\subsection{Nonsingular Universe Construction}

A natural approach to resolving the singularity problem of general
relativity is to consider an effective theory of gravity which
contains higher order terms, in addition to the Ricci scalar
of the Einstein action. This approach is well motivated, since
any effective action for classical gravity obtained
from string theory, quantum gravity, or by integrating out matter
fields, will contain higher derivative terms. Thus, it is quite natural
to consider higher derivative effective gravity theories when studying
the properties of space-time at large curvatures.

Most higher derivative gravity theories have much worse singularity problems than Einstein's theory. However, it is not unreasonable to expect that in the fundamental theory of nature, be it string theory or some other theory, the curvature of space-time is limited. In Ref. \cite{Markov} the hypothesis was made that when the limiting curvature is reached, the geometry must approach that of a maximally symmetric space-time, namely de Sitter space. The question now becomes whether it is possible to find a class of higher derivative effective actions for gravity which have the property that at large curvatures the solutions approach de Sitter space. A {\it nonsingular Universe construction} which achieves this goal was proposed in Refs. \cite{MB92,BMS93}. It is based on adding to the Einstein action a particular combination of quadratic invariants of the Riemann tensor chosen such that the invariant vanishes only in de Sitter space-times. This invariant is coupled to the Einstein action via a Lagrange multiplier field in a way that the Lagrange multiplier constraint equation forces the invariant to zero at high curvatures. Thus, the metric becomes de Sitter and hence explicitly nonsingular.
  
If successful, the above construction will have some very appealing 
consequences.  Consider, for example, a collapsing spatially 
homogeneous Universe.  According to Einstein's theory, this Universe 
will collapse in a finite proper time to a final ``big crunch" singularity.
In the new theory, however, the Universe will approach a de Sitter model as 
the curvature increases. If the 
Universe is closed, there will be a de Sitter bounce followed by 
re-expansion.  Similarly, spherically 
symmetric vacuum solutions of the new equations of motion will presumably be nonsingular, i.e., black holes 
would have no singularities in their centers. This 
would have interesting consequences for the black hole information 
loss problem. In two dimensions, this construction has been successfully realized \cite{TMB93}.

The {\it nonsingular Universe construction} of \cite{MB92,BMS93} and its applications to dilaton cosmology \cite{BEM98,EB99} are reviewed in an accompanying article in these proceedings \cite{BM99}. Here is just a very
brief summary of the points relevant to the problems listed in Section 5.    

The procedure for obtaining a nonsingular Universe theory \cite{MB92} is based 
on a Lagrange multiplier construction.  
Starting from the Einstein action, one can introduce a Lagrange 
multiplier $\varphi_1$ coupled to the Ricci scalar $R$ to obtain a 
theory with bounded $R$:
\be \label{act1}
S = \int d^4 x \sqrt{-g} (R + \varphi_1 \, R + V_1 (\varphi_1) ) \, , 
\ee
where the potential $V_1 (\varphi_1)$ satisfies the asymptotic 
conditions coming from demanding that at small values of $\varphi_1$ (small curvature), the Einstein theory is recovered, and that at large values of $\varphi_1$ the Ricci scalar tends to a constant.

However, this action is insufficient to obtain a nonsingular gravity 
theory.  For example, singular solutions of the Einstein equations 
with $R=0$ are not affected at all.  The minimal requirements for a 
nonsingular theory are that {\it all} curvature invariants remain 
bounded and the space-time manifold is geodesically complete.  
It is possible to achieve this by a two-step procedure.  First, we choose one 
curvature invariant $I_1 (g_{\mu\nu})$ (e.g. $I_1 = R$ in (\ref{act1})) and demand that it be 
explicitely bounded by the construction of (\ref{act1}).  In a second step, we demand that as $I_1 (g_{\mu\nu})$ approaches its limiting value, the metric $g_{\mu\nu}$ approach 
the de Sitter metric $g^{DS}_{\mu\nu}$, a definite nonsingular metric 
with maximal symmetry.  In this case, all curvature invariants are 
automatically bounded (they approach their de Sitter values), and the 
space-time can be extended to be geodesically complete.
The second step can be implemented by 
another Lagrange multiplier construction \cite{MB92}.  Consider a curvature 
invariant $I_2 (g_{\mu\nu})$ with the property that 
\be
I_2 (g_{\mu\nu}) = 0 \>\> \Leftrightarrow \>\> g_{\mu\nu} = 
g^{DS}_{\mu\nu} \, .
\ee
Next, introduce a second Lagrange multiplier field $\varphi_2$ which couples 
to $I_2$ and choose a potential $V_2 (\varphi_2)$ which forces $I_2$ 
to zero at large $|\varphi_2|$:
\be
S = \int d^4  x \sqrt{-g} [ R + \varphi_1 I_1 + V_1 (\varphi_1) + 
\varphi_2 I_2 + V_2 (\varphi_2) ] \, , 
\ee
with asymptotic conditions  
\begin{eqnarray}
V_2 (\varphi_2) & \sim &{\rm const} \>\> {\rm as} \> | 
\varphi_2 | \rightarrow \infty \label{asympt3} \\
V_2 (\varphi_2) & \sim & \varphi^2_2 \>\> {\rm as} \> |\varphi_2 | 
\rightarrow 0 \, , \label{asympt4}
\end{eqnarray}
for $V_2 (\varphi_2)$.  The first constraint forces $I_2$ to zero, the 
second is required in order to obtain the correct low curvature limit.

The invariant  
\be \label{inv}
I_2 = (4  R_{\mu\nu} R^{\mu\nu} - R^2 + C^2)^{1/2} \, ,
\ee
singles out the de Sitter metric among all homogeneous and isotropic 
metrics (in which case adding $C^2$, the Weyl tensor square, is 
superfluous), all homogeneous and anisotropic metrics, and all 
radially symmetric metrics.

As a specific example one can consider the action \cite{MB92,BMS93}
\be \label{act2}
S = \int d^4 x \sqrt{-g} \left[ R + \varphi_1 R - (\varphi_2 + 
{3\over{\sqrt{2}}} \varphi_1) I_2^{1/2} + V_1 (\varphi_1) + V_2 
(\varphi_2) \right] 
\ee
with
\begin{eqnarray}
V_1 (\varphi_1) & = & 12 \, H^2_0 {\varphi^2_1\over{1 + \varphi_1}} \left( 1 
- {\ln (1 + \varphi_1)\over{1 + \varphi_1}} \right) \\
V_2 (\varphi_2) & =  & - 2 \sqrt{3} \, H^2_0 \, {\varphi^2_2\over{1 + 
\varphi^2_2}} \, .
\end{eqnarray}
It can be shown that all solutions of the equations of motion which follow from this action are nonsingular. They are either periodic about Minkowski space-time $(\varphi_1, \varphi_2) = (0, 0)$ or else asymptotically approach de Sitter space ($|\varphi_2 | \rightarrow \infty$).

One of the most interesting properties of this theory is asymptotic 
freedom \cite{BMS93}, i.e., the coupling between matter and gravity goes to 
zero at high curvatures.  It is easy to add matter (e.g., dust, 
radiation or a scalar field) to the gravitational action in the standard way.
One finds that in the asymptotic de Sitter regions, the trajectories of 
the solutions projected onto the $(\varphi_1, \, \varphi_2)$ plane are 
unchanged by adding matter.  This applies, for example, in a phase of de Sitter 
contraction when the matter energy density is increasing exponentially 
but does not affect the metric.  The physical reason for asymptotic 
freedom is obvious: in the asymptotic regions of phase space, the 
space-time curvature approaches its maximal value and thus cannot be 
changed even by adding an arbitrarily high matter energy density.
Hence, there is the possibility that this theory will admit a natural suppression mechanism for cosmological fluctuations. If this were the case, then the solution of the singularity problem would simultaneously help resolve the fluctuation problem of potential-driven inflationary cosmology.
 
\subsection{Back-Reaction of Cosmological Perturbations}

The linear theory of cosmological perturbations in inflationary cosmology is well studied. However, effects beyond linear order have received very little attention. Beyond linear order, perturbations can effect the background in which they propagate, an effect well known from early studies \cite{Brill} of gravitational waves. As will be summarized below, the back-reaction of cosmological perturbations in an exponentially expanding Universe acts like a negative cosmological constant, as first realized in the context of studies of gravitational waves in de Sitter space in \cite{TW}.

Gravitational back-reaction of cosmological perturbations concerns itself with the evolution of space-times which consist of small fluctuations about a symmetric Friedmann-Robertson-Walker space-time with metric $g_{\mu \nu}^{(0)}$. The goal is to study the evolution of spatial averages of observables in the perturbed space-time. In linear theory, such averaged quantities evolve like the corresponding variables in the background space-time. However, beyond linear theory perturbations have an effect on the averaged quantities. In the case of gravitational waves, this effect is well known \cite{Brill}: gravitational waves carry energy and momentum which affect the background in which they propagate. Here, we shall focus on scalar metric perturbations.

The analysis of gravitational back-reaction \cite{ABM1} is related to early work by Brill, Hartle and Isaacson \cite{Brill}, among others. The idea is to expand the Einstein equations to second order in the perturbations, to assume that the first order terms satisfy the equations of motion for linearized cosmological perturbations \cite{MFB92} (hence these terms cancel), to take the spatial average of the remaining terms, and to regard the resulting equations as equations for a new homogeneous metric $g_{\mu \nu}^{(0, br)}$ which includes the effect of the perturbations to quadratic order:
\be \label{breq}
G_{\mu \nu}(g_{\alpha \beta}^{(0, br)}) \, = \, 8 \pi G \left[ T_{\mu \nu}^{(0)} + \tau_{\mu \nu} \right]\,
\ee
where the effective energy-momentum tensor $\tau_{\mu \nu}$ of gravitational back-reaction contains the terms resulting from spatial averaging of the second order metric and matter perturbations:
\be \label{efftmunu}
\tau_{\mu \nu} \, = \, < T_{\mu \nu}^{(2)} - {1 \over {8 \pi G}} G_{\mu \nu}^{(2)} > \, ,
\ee
where pointed brackets stand for spatial averaging, and the superscripts indicate the order in perturbations.

As formulated in (\ref{breq}) and (\ref{efftmunu}), the back-reaction problem is not independent of the choice of coordinates in space-time and hence is not well defined. It is possible to take a homogeneous and isotropic space-time, choose different coordinates, and obtain a nonvanishing $\tau_{\mu \nu}$. This ``gauge" problem is related to the fact that in the above prescription, the hypersurface over which the average is taken depends on the choice of coordinates. 

The key to resolving the gauge problem is to realize that to second order in perturbations, the background variables chage. A gauge independent form of the back-reaction equation (\ref{breq}) can hence be derived \cite{ABM1} by defining background and perturbation variables $Q = Q^{(0)} + \delta Q$ which do not change under linear coordinate transformations. Here, $Q$ represents collectively both metric and matter variables. The gauge-invariant form of the back-reaction equation then looks formally identical to (\ref{breq}), except that all variables are replaced by the corresponding gauge-invariant ones. We will follow the notation of \cite{MFB92}, and use as gauge-invariant perturbation variables the Bardeen potentials \cite{Bardeen} $\phi$ and $\Psi$ which in longitudinal gauge coincide with the actual metric perturbations $\delta g_{\mu \nu}$. Calculations hence simplify greatly if we work directly in  longitudinal gauge. Recently, these calculations have been confirmed \cite{WA} by working in a completely different gauge, making use of the covariant approach.
 
In \cite{ABM2}, the effective energy-momentum tensor $\tau_{\mu \nu}$ of gravitational back-reaction was evaluated for long wavelength fluctuations in an inflationary Universe in which the matter responsible for inflation is a scalar field $\varphi$ with the potential
\be
V(\varphi) \, = \, {1 \over 2} m^2 \varphi^2 \, .
\ee
Since there is no anisotropic stress in this model, in longitudinal gauge the perturbed metric can be written \cite{MFB92} in terms of a
single gravitational potential $\phi$
\be
ds^2 =  (1+ 2 \phi) dt^2 - a(t)^2(1 - 2\phi) \delta_{i j} dx^i dx^j  \, ,
\ee
where $a(t)$ is the cosmological scale factor. 

It is now straightforward to compute $G_{\mu \nu}^{(2)}$ and 
$T_ {\mu \nu}^{(2)}$ in terms of the background fields and the metric and matter fluctuations $\phi$ and $\delta \varphi$, By taking averages and making use of (\ref{efftmunu}), the effective energy-momentum tensor $\tau_{\mu \nu}$ can be computed \cite{ABM2}.

The general expressions for the effective energy density $\rho^{(2)} = \tau^0_0$ and effective pressure $p^{(2)} = - {1 \over 3} \tau^i_i$ involve many terms. However, they greatly simplify if we consider perturbations with wavelength greater than the Hubble radius. In this case, all terms involving spatial gradients are negligible. From the theory of linear cosmological perturbations (see e.g. \cite{MFB92}) it follows that on scales larger than the Hubble radius the time derivative of $\phi$ is also negligible as long as the equation of state of the background does not change. The Einstein constraint equations relate the two perturbation variables $\phi$ and $\delta \varphi$, enabling scalar metric and matter fluctuations to be described in terms of a single gauge-invariant potential $\phi$. During the slow-rolling period of the inflationary Universe, the constraint equation takes on a very simple form and implies that $\phi$ and $\delta \varphi$ are proportional. The upshot of these considerations is that $\tau_{\mu \nu}$ is proportional to the two point function $< \phi^2 >$, with a coefficient tensor which depends on the background dynamics. In the slow-rolling approximation we obtain \cite{ABM2}
\be
\rho^{(2)} \, \simeq \, - 4 V < \phi^2 >
\ee
and
\be
p^{(2)} \, = \, - \rho^{(2)} \, .
\ee
This demonstrates that the effective energy-momentum tensor of long-wavelength cosmological perturbations has the same form as a negative cosmological constant. This back-reaction mechanism may thus relate closely to the cosmological constant problem \cite{RBTexas}.

\section{Conclusions}

Inflationary cosmology is an attractive {\it scenario}. It solves some problems of standard cosmology and leads to the possibility of a causal theory of structure formation. The specific predictions of an inflationary model of structure formation, however, depend on the specific realization of inflation, which makes the idea of inflation hard to verify or falsify. Many models of inflation have been suggested, but at the present time none are sufficiently distinguished to form a ``standard" inflationary theory.

There has been a lot of recent progress in inflationary cosmology. As explained in Section 4.1, a new theory of inflationary reheating (preheating) has been developed based on parametric resonance. Preheating has dramatic consequences for baryogenesis and for the production of particles and solitons at the end of inflation. A consistent quantum theory of the generation and evolution of linear cosmological perturbations has been developed (see Section 4.2). At the present time, a lot of work is being devoted to extend the analysis of cosmological perturbations beyond linear order. A third area of dramatic progress in inflationary cosmology has been the development of precision calculations of the power spectrum of density fluctuations and of CMB anisotropies which will allow detailed comparisons between current and upcoming observations and inflationary models.

However, there are important unsolved problems of principle in inflationary cosmology. Four such problems discussed in these lectures (in Section 5) are the {\it fluctuation problem}, the {\it super-Planck-scale physics problem}, the {\it singularity problem} and the {\it cosmological constant problem}, the last of which is probably the Achilles heel of inflationary cosmology. 

It may be that a convincing realization of inflation will have to wait for an improvement in our understanding of fundamental physics. In Section 6, we described some promising but incomplete avenues which address some of the above problems, while still yielding an inflationary epoch.


\begin{thebibliography}{10}  

\bibitem{Guth} {A. Guth, {\it Phys. Rev.} {\bf D23}, 347 (1981).}
\bibitem{Linde} {A. Linde, `Particle Physics and Inflationary Cosmology'
(Harwood, Chur, 1990).}
\bibitem{GuthBlau} S. Blau and A. Guth, `Inflationary Cosmology,' in `300 Years
of Gravitation' ed. by S. Hawking and W. Israel (Cambridge Univ.
Press, Cambridge, 1987).
\bibitem{Olive} K. Olive, {\it Phys. Rep.} {\bf 190}, 307 (1990).
\bibitem{LR99} D. Lyth and A. Riotto, {\it Phys. Rept.} {\bf 314}, 1 (1999).
\bibitem{Liddle2} A. Liddle, ``An Introduction to Cosmological Inflation", astro-ph/9901124.
\bibitem{Peebles} P.J.E. Peebles, `Principles of Physical Cosmology' (Princeton Univ. Press, Princeton, 1993).
\bibitem{Padmanabhan} T. Padmanabhan, `Structure Formation in the Universe' (Cambridge Univ. Press, Cambridge, 1993).
\bibitem{Peacock} J. Peacock, `Cosmological Physics' (Cambridge Univ. Press, Cambridge, 1999).
\bibitem{CFA} {V. de Lapparent, M. Geller and J. Huchra, {\it Ap. J.
(Lett)} {\bf 302}, L1 (1986).}
\bibitem{LCRS} S. Landy, S. Shectman, H. Lin, R. Kirschner, A. Oemler and D. Tucker, {\it Ap. J. (Lett.)} {\bf 456}, 1L (1996);\\
H. Lin, R. Kirshner, S. Shectman, S. Landy, A. Oemler, D. Tucker and P. Schechter, {\it Ap. J. (Lett.)} {\bf 464}, 60L (1996);\\
S. Shectman, S. Landy, A. Oemler, D. Tucker, H. Lin, R. Kirshner and P. Schechter, {\it Ap. J. (Suppl.)} {\bf 470}, 172S (1996).
\bibitem{explosion} {J. Ostriker and L. Cowie, {\it Ap. J. (Lett.)} {\bf
243}, L127 (1981).}
\bibitem{VilShell} A. Vilenkin and E.P.S. Shellard, `Strings and Other Topological Defects' (Cambridge Univ. Press, Cambridge, 1994).
\bibitem{HK95} M. Hindmarsh and T.W.B. Kibble, {\it Rept. Prog. Phys.} {\bf 58}, 477 (1995).
\bibitem{RB94} R. Brandenberger, {\it Int. J. Mod. Phys.} {\bf A9}, 2117 (1994).
\bibitem{MB99} J. Magueijo and R. Brandenberger, these proceedings.
\bibitem{RB85} {R. Brandenberger, {\it Rev. Mod. Phys.} {\bf 57}, 1
(1985).}
\bibitem{Kazanas} {D. Kazanas, {\it Ap. J.} {\bf 241}, L59 (1980).}
\bibitem{open} J. Gott III and T. Statler, {\it Phys. Lett.} {\bf 136B}, 157 (1984);\\
M. Bucher, A. Goldhaber and N. Turok, {\it Phys. Rev.} {\bf D52}, 3314 (1995);\\
A. Linde, {\it Phys. Lett.} {\bf B351}, 99 (1995).
\bibitem{Press} {W. Press, {\it Phys. Scr.} {\bf 21}, 702 (1980).}
\bibitem{Mukh80} {G. Chibisov and V. Mukhanov, `Galaxy Formation and
Phonons,' Lebedev Physical Institute Preprint No. 162 (1980);
\\
G. Chibisov and V. Mukhanov, {\it Mon. Not. R. Astron. Soc.} {\bf
200}, 535 (1982).}
\bibitem{Lukash} {V. Lukash, {\it Pis'ma Zh. Eksp. Teor. Fiz.} {\bf 31}, 631
(1980).}
\bibitem{Sato} {K. Sato, {\it Mon. Not. R. Astron. Soc.} {\bf 195},
467 (1981).}
\bibitem{RB83} R. Brandenberger, {\it Phys. Lett.} {\bf 129B}, 397 (1983).
\bibitem{GuthTye} {A. Guth and S.-H. Tye, {\it Phys. Rev. Lett.} {\bf
44}, 631 (1980).}
\bibitem{finiteT} {D. Kirzhnits and A. Linde, {\it Pis'ma Zh. Eksp.
Teor. Fiz.} {\bf 15}, 745 (1972); \\
D. Kirzhnits and A. Linde, {\it Zh. Eksp. Teor. Fiz.} {\bf 67}, 1263
(1974);\\
C. Bernard, {\it Phys. Rev.} {\bf D9}, 3313 (1974);\\
L. Dolan and R. Jackiw, {\it Phys. Rev.} {\bf D9}, 3320 (1974);\\
S. Weinberg, {\it Phys. Rev.} {\bf D9}, 3357 (1974).}
\bibitem{decay} {S. Coleman, {\it Phys. Rev.} {\bf D15}, 2929 (1977);\\
C. Callan and S. Coleman, {\it Phys. Rev.} {\bf D16}, 1762 (1977);\\
M. Voloshin, Yu. Kobzarev and L. Okun, {\it Sov. J. Nucl. Phys.} {\bf 20}, 644 (1975);\\
M. Stone, {\it Phys. Rev.} {\bf D14}, 3568 (1976);\\
M. Stone, {\it Phys. Lett.} {\bf 67B}, 186 (1977);\\
P. Frampton, {\it Phys. Rev. Lett.}, {\bf 37}, 1380 (1976);\\
S. Coleman, in `The Whys of Subnuclear Physics' (Erice 1977), ed
by A. Zichichi (Plenum, New York, 1979).}
\bibitem{GuthWein} {A. Guth and E. Weinberg, {\it Nucl. Phys.} {\bf B212}, 321 (1983).}
\bibitem{MJ84} C. Mathiazhagan and V. Johri, {\it Class. Quant. Grav.} {\bf 1}, L29 (1984).
\bibitem{LS89} D. La and P. Steinhardt, {\it Phys. Rev. Lett.} {\bf 62}, 376 (1989).
\bibitem{Linde82} {A. Linde, {\it Phys. Lett.} {\bf 108B}, 389 (1982).}
\bibitem{AS82}
{A. Albrecht and P. Steinhardt, {\it Phys. Rev. Lett.} {\bf 48}, 1220
(1982).}
\bibitem{Linde83} {A. Linde, {\it Phys. Lett.} {\bf 129B}, 177 (1983).}
\bibitem{Linde94} {A. Linde, D. Linde and A. Mezhlumian, {\it Phys. Rev.} {\bf D49}, 1783 (1994); \\
A. Linde, `Lectures on Inflationary Cosmology', Stanford preprint SU-ITP-94-36,
hep-th/9410082 (1994).}
\bibitem{Starob87} {A. Starobinski, in `Current Trends in Field Theory, Quantum
Gravity, and Strings', Lecture Notes in Physics, ed. by H. de Vega and N.
Sanchez (Springer, Heidelberg, 1986).}
\bibitem{ES95} E. Stewart, {\it Phys. Rev.} {\bf D51}, 6847 (1995).
\bibitem{ARS96} J. Adams, G. Ross and S. Sarkar, {\it Phys. Lett.} {\bf B391}, 271 (1997).
\bibitem{hybrid} A. Linde, {\it Phys. Rev.} {\bf D49}, 748 (1994).
\bibitem{hybridSG} P. Binetruy and G. Dvali, {\it Phys. Lett.} {\bf D388}, 241 (1996);\\
E. Halyo, {\it Phys. Lett.} {\bf B387}, 43 (1996);\\
A. Linde and A. Riotto, {\it Phys. Rev.} {\bf D56}, R1844 (1997).
\bibitem{Banks} T. Banks, `Remarks on M Theoretic Cosmology', hep-th/9906126.
\bibitem{HW} P. Horava and E. Witten, {\it Nucl. Phys.} {\bf B460}, 506 (1996).
\bibitem{braneinfl} G. Dvali and S. Tye, {\it Phys. Lett.} {\bf B450}, 72 (1999);\\
N. Kaloper and A. Linde, {\it Phys. Rev.} {\bf D59}, 101303 (1999).
\bibitem{AS80} {A. Starobinsky, {\it Phys. Lett.} {\bf 91B}, 99 (1980).}
\bibitem{PBB} M. Gasperini and G. Veneziano, {\it Astropart. Phys.} {\bf 1}, 317 (1993).
\bibitem{DM98} T. Damour and V. Mukhanov, {\it Phys. Rev. Lett.} {\bf 80}, 3440 (1998).
\bibitem{Mukh81} V. Mukhanov and G. Chibisov, {\it JETP Lett.} {\bf 33}, 532 (1981).
\bibitem{flucts} A. Guth and S.-Y. Pi, {\it Phys. Rev. Lett.} {\bf 49}, 110 (1982);\\ 
S. Hawking, {\it Phys. Lett.} {\bf 115B}, 295 (1982);\\ 
A. Starobinski, {\it Phys. Lett.} {\bf 117B}, 175 (1982);\\ 
V. Mukhanov, {\it JETP Lett.} {\bf 41}, 493 (1985).
\bibitem{BST} J. Bardeen, P. Steinhardt and M. Turner, {\it Phys. Rev.} {\bf D28}, 1809 (1983).
\bibitem{SW} R. Sachs and A. Wolfe, {\it Astrophys. J.} {\bf 147}, 73 (1967).
\bibitem{HZ} {E. Harrison, {\it Phys. Rev.} {\bf D1}, 2726 (1970); \\
Ya.B. Zel'dovich, {\it Mon. Not. R. astron. Soc.} {\bf 160}, 1p
(1972).}
\bibitem{APM} C. Baugh and G. Efstathiou, {\it Mon. Not. R. astr. Soc.} {\bf 267}, 323 (1994).
\bibitem{LCRSPS} H. Lin et al., {\it Astrophys. J.} {\bf 471}, 617 (1996).
\bibitem{COBE} G. Smoot et al., {\it Astrophys. J. (Lett.)} {\bf 396}, L1 (1992).
\bibitem{KLS94} {L. Kofman, A. Linde and A. Starobinski, {\it Phys. Rev. Lett.} {\bf 73}, 3195 (1994).}
\bibitem{STB95} Y. Shtanov, J. Traschen and R. Brandenberger, {\it Phys. Rev.} {\bf D51}, 5438 (1995).
\bibitem{DolLin} A. Dolgov and A. Linde, {\it Phys. Lett.} {\bf 116B}, 329 (1982).
\bibitem{AFW} {L. Abbott, E. Farhi and M. Wise, {\it Phys. Lett.} {\bf 117B}, 29
(1982).}
\bibitem{TB90} {J. Traschen and R. Brandenberger, {\it Phys. Rev.} {\bf D42},
2491 (1990).}
\bibitem{parres} {L. Landau and E. Lifshitz, `Mechanics' (Pergamon, Oxford,
1960); \\
V. Arnold, `Mathematical Methods of Classical Mechanics' (Springer,
New York, 1978).}
\bibitem{KLS97} {L. Kofman, A. Linde and A. Starobinski, {\it Phys. Rev.} {\bf D56}, 3258 (1997).}
\bibitem{Yoshimura2} M. Hotta, I. Joichi, S. Matsumoto and M. Yoshimura, {\it Phys. Rev.} {\bf D55}, 4614 (1997).
\bibitem{Birrell} {N. Birrell and P. Davies, `Quantum Fields in Curved Space'
(Cambridge Univ. Press, Cambridge, 1982).}
\bibitem{KLS96} L. Kofman, A. Linde and A. Starobinski, {\it Phys. Rev. Lett.} {\bf 76}, 1011 (1996).
\bibitem{Tkachev} I. Tkachev, {\it Phys. Lett.} {\bf B376}, 35 (1996).
\bibitem{KLR96} E. Kolb, A. Linde and A. Riotto, {\it Phys. Rev. Lett.} {\bf 77}, 4290 (1996).
\bibitem{therm} S. Khlebnikov and I. Tkkachev, {\it Phys. Rev. Lett.} {\bf 77}, 219 (1996);\\
S. Khlebnikov and I. Tkachev, {\it Phys. Lett.} {\bf B390}, 80 (1997);\\
T. Prokopec and T. Roos, {\it Phys. Rev.} {\bf D55}, 3768 (1997);\\
B. Greene, T. Prokopec and T. Roos, {\it Phys. Rev.} {\bf D56}, 6484 (1997).
\bibitem{KK97} S. Kasuya and M. Kawasaki, {\it Phys. Rev.} {\bf D56}, 7597 (1997);\\
S. Kasuya and M. Kawasaki, {\it Phys. Rev.} {\bf D58}, 083516 (1998).
\bibitem{PS98} M. Parry and A. Sornborger, `Domain wall production during inflationary reheating', hep-ph/9805211.
\bibitem{BKM1} B. Bassett, D. Kaiser and R. Maartens, {\it Phys. Lett.} {\bf B455}, 84 (1999).
\bibitem{FB99} F. Finelli and R. Brandenberger, {\it Phys. Rev. Lett.} {\bf 82}, 1362 (1999).
\bibitem{PE99} M. Parry and R. Easther, {\it Phys. Rev.} {\bf D59}, 061301 (1999).
\bibitem{JS99} K. Jedamzik and G. Sigl, `On metric preheating', hep-ph/9906287.
\bibitem{PI99} P. Ivanov, `On generation of metric perturbations during preheating', astro-ph/9906415.
\bibitem{BKM2} B. Bassett, F. Tamburini, D. Kaiser and R. Maartens, `Metric preheating and limitations of linearized gravity 2', hep-ph/9901319. 
\bibitem{Lifshitz} {E. Lifshitz, {\it Zh. Eksp. Teor. Fiz.} {\bf 16}, 587 (1946); \\
E. Lifshitz and I. Khalatnikov, {\it Adv. Phys.} {\bf 12}, 185 (1963).}
\bibitem{PressVish} {W. Press and E. Vishniac, {\it Ap. J.} {\bf 239}, 1 (1980).}
\bibitem{MFB92} {V. Mukhanov, H. Feldman and R. Brandenberger, {\it Phys.
Rep.} {\bf 215}, 203 (1992).}
\bibitem{MFB92} {R. Brandenberger, H. Feldman, V. Mukhanov and T. Prokopec,
`Gauge Invariant Cosmological Perturbations: Theory and Applications,'
publ. in ``The Origin of Structure in the Universe," eds. E. Gunzig and P.
Nardone (Kluwer, Dordrecht, 1993).}
\bibitem{Bardeen} {J. Bardeen, {\it Phys. Rev.} {\bf D22}, 1882 (1980).}
\bibitem{BKP83} {R. Brandenberger, R. Kahn and W. Press, {\it Phys. Rev.} {\bf
D28}, 1809 (1983).}
\bibitem{KoSa84} {H. Kodama and M. Sasaki, {\it Prog. Theor. Phys. Suppl.} No.
78, 1 (1984).}
\bibitem{Durrer} {R. Durrer and N. Straumann, {\it Helvet. Phys. Acta} {\bf
61}, 1027 (1988);\\
R. Durrer, {\it Fund. Cosmic Phys.} {\bf 15}, 209 (1994), astro-ph/9311040.}
\bibitem{Lyth} {D. Lyth, {\it Phys. Rev.} {\bf D31}, 1792 (1985);\\
D. Lyth and M. Mukherjee, {\it Phys. Rev.} {\bf D38}, 485
(1988).}
\bibitem{Hwang} J. Hwang and E. Vishniac, {\it Ap. J.} {\bf 353}, 1 (1990).
\bibitem{EllisBruni} {G.F.R. Ellis and M. Bruni, {\it Phys. Rev.} {\bf D40}, 1804 (1989);\\
G.F.R. Ellis, J. Hwang and M. Bruni, {\it Phys. Rev.} {\bf D40}, 1819 (1989).}
\bibitem{Salopek} D. Salopek and J. Stewart, {\it Phys. Rev.} {\bf D51}, 517 (1995).
\bibitem{Stewart} {J. Stewart, {\it Class. Quantum Grav.} {\bf 7}, 1169 (1990).}
\bibitem{SteWa} {J. Stewart and M. Walker, {\it Proc. R. Soc.} {\bf A341}, 49
(1974).}
\bibitem{BK84} R. Brandenberger and R. Kahn, {\it Phys. Rev.} {\bf D28}, 2172 (1984).
\bibitem{RBrev} {R. Brandenberger, ``Modern Cosmology and Structure Formation", in `CP Violation and the Limits of the Standard Model (TASI 94)', ed. J. Donoghue (World Scientific, Singapore, 1995), astro-ph/9411049;\\
R. Brandenberger, in `Physics of the Early Universe,' proc.
of the 1989 Scottish Univ. Summer School in Physics, ed. by J. Peacock, A.
Heavens and A. Davies (SUSSP Publ., Edinburgh, 1990);\\
R. Brandenberger, `Lectures on Modern Cosmology and Structure Formation',  in  
`Particles and Fields', ed. by O. Eboli and V. Ribelles (World Scientific, Singapore 1994).} 
\bibitem{Gotz} J. Martin and D. Schwarz, {\it Phys. Rev.} {\bf D57}, 3302 (1998);\\
M. Gotz, {\it Mon. Not. Roy. Astron. Soc.} {\bf 295}, 873 (1998), astro-ph/9704271.
\bibitem{Mukh88} V. Mukhanov, {\it Zh. Eksp. Teor. Fiz.} {\bf 94}, 1 (1988).
\bibitem{Sasaki} M. Sasaki, {\it Prog. Theor. Phys.} {\bf 76}, 1036 (1986).
\bibitem{Freese} {F. Adams, K. Freese and A. Guth, {\it Phys. Rev.} {\bf D43},
965 (1991).}
\bibitem{Jacobson} T. Jacobson, `Black Hole Evaporation: An Open Question', in
{\it Brighton 1990, Relativistic Astrophysics, Cosmology, and Fundamental Physics}, Proc. of the 1990 TEXAS/ESO-CERN symposium, ed. by J. Barrow, L. Mestel and P. Thomas (New York Acad. Sci., New York, 1991).
\bibitem{PBBflucts} R. Brustein, M. Gasperini, M. Giovannini, V. Mukhanov and G. Veneziano, {\it Phys. Rev.} {\bf D51}, 6744 (1995).
\bibitem{PBBaxion} E. Copeland, R. Easther and D. Wands, {\it Phys. Rev.} {\bf D56}, 874 (1997).
\bibitem{Borde} A. Borde and A. Vilenkin, {\it Phys. Rev. Lett.} {\bf 72},  3305  (1993).
\bibitem{cosmorev} {S. Weinberg, {\it Rev. Mod. Phys.} {\bf 61}, 1 (1989);\\
S. Carroll, W. Press and E. Turner, {\it Ann. Rev. Astron. Astrophys.} {\bf
30}, 499 (1992).}
\bibitem{PBBEP} R. Brustein and G. Veneziano, {\it Phys. Lett.} {\bf B329}, 429 (1994);\\
R. Easther, K. Maeda and D. Wands, {\it Phys. Rev.} {\bf D53}, 4247 (1996);\\
N. Kaloper, R. Madden and K. Olive, {\it Phys. Lett.} {\bf B371}, 34 (1995).
\bibitem{PBBIC} N. Kaloper, A. Linde and R. Bousso, {\it Phys. Rev.} {\bf D59}, 043508 (1999).
\bibitem{BDV} A. Buonanno, T. Damour and G. Veneziano, {\it Nucl. Phys.} {\bf B543}, 275 (1999).
\bibitem{BZ97} R. Brandenberger and A. Zhitnitsky, {\it Phys. Rev.} {\bf D55}, 4640 (1997).
\bibitem{BM92} R. Ball and A. Matheson, {\it Phys. Rev.} {\bf D45}, 2647 (1992);\\
A. Matheson, R. Ball, A.-C. Davis and R. Brandenberger, {\it Nucl. Phys.} {\bf B328}, 223 (1989).
\bibitem{ARZ} A. Zhitnitsky, {\it Phys. Rev.} {\bf D54}, 5148 (1996).
\bibitem{AL82b} A. Linde, {\it Phys. Lett.} {\bf 116B}, 335 (1982).
\bibitem{VilFord} A. Vilenkin and L. Ford, {\it Phys. Rev.} {\bf D25}, 2569 (1982).
\bibitem{Markov} M. Markov, {\it Pis'ma Zh. Eksp. Theor. Fiz.} {\bf 36}, 214 
(1982); \\
M. Markov, {\it Pis'ma Zh. Eksp. Theor. Fiz.} {\bf 46}, 342 (1987); 
\\
V. Ginsburg, V. Mukhanov and V. Frolov,  {\it Pis'ma Zh. Eksp. Theor. Fiz.} 
{\bf 94}, 3 (1988); \\
V. Frolov, M. Markov and V. Mukhanov, {\it Phys. Rev.} {\bf D41}, 383 
(1990).
\bibitem{MB92} V. Mukhanov and R. Brandenberger, {\it Phys. Rev.
Lett.} {\bf 68}, 1969 (1992).
\bibitem{BMS93} R. Brandenberger, V. Mukhanov and A. Sornborger, {\it Phys. Rev.} {\bf D48}, 1629 (1993).
\bibitem{TMB93} M. Trodden, V. Mukhanov and R. Brandenberger, {\it Phys. 
Lett.} {\bf B316}, 483 (1993).
\bibitem{BEM98} R. Brandenberger, R. Easther and J. Maia, {\it JHEP} {\bf 9808}, 7 (1998), gr-qc/9806111.
\bibitem{EB99} D. Easson and R. Brandenberger, {\it JHEP} {\bf 9909} 3 (1999), hep-th/9905175.
\bibitem{BM99} R. Brandenberger and J. Magueijo, `Imaginative cosmology', these proceedings.
\bibitem{Brill} D. Brill and J. Hartle, {\it Phys. Rev.} {\bf 135} 
B271-278 (1964);
R. Isaacson, {\it Phys. Rev.} {\bf 166}, 1263 and 1272 (1968).
\bibitem{TW} N. Tsamis and R. Woodard, {\it Phys. Lett.} {\bf B292}, 269 (1993); {\it Phys. Rev.} {\bf D54}, 2621 (1996); {\it Nucl. Phys.} {\bf B474}, 235 (1996); {\it Ann. Phys.} {\bf 253}, 1 (1997).
\bibitem{ABM1} V. Mukhanov, L.R.W. Abramo and R. Brandenberger, {\it Phys. Rev. Lett.} {\bf 78}, 1624 (1997).
\bibitem{WA} L.R.W. Abramo and R. Woodard, {\it Phys. Rev.} {\bf D60}, 044010 (1999), astro-ph/9811430.
\bibitem{ABM2} L.R.W. Abramo, R. Brandenberger and V. Mukhanov, {\it Phys. Rev.} {\bf D56}, 3248 (1997), gr-qc/9704037.
\bibitem{RBTexas} R. Brandenberger, contribution to the 19th Texas Symposium on Relativistic Astrophysics, Paris, France, Dec. 14 - 18 1998, to be publ. in the proceedings (CD-ROM, {\it Nucl. Phys. B, Suppl.}), Brown preprint BROWN-HET-1180.


\end{thebibliography}
\end{document}